\documentclass[aps,prd,article,superscriptaddress,preprintnumbers,nofootinbib]{revtex4-2}
\usepackage{hyphenat}
\usepackage{comment}
\usepackage[utf8]{inputenc} 
\usepackage{graphicx,color,overpic,mathtools}
\usepackage{amsthm,amsmath,amssymb,hyperref,mathrsfs}

\usepackage[dvipsnames, table, svgnames]{xcolor}
\usepackage{diagbox}
\usepackage{makecell}
\usepackage{stackengine}
\usepackage{adjustbox}
\usepackage{amsmath}
\usepackage{nicematrix}

\usepackage{mathrsfs,bm}
\usepackage{natbib,textcase}
\usepackage{flexisym}
\usepackage{braket,bm,bbm,setspace}
\PassOptionsToPackage{normalem}{ulem}
\usepackage{ulem} 
\usepackage{physics}
\usepackage{float}
\usepackage[makeroom]{cancel}
\usepackage[english]{babel}
\usepackage{graphicx}
\graphicspath{ {images/} }
\usepackage{breqn}
\usepackage{bm}
\addto\captionsspanish{}
\hypersetup{
    colorlinks=false,
    pdfborder={0 0 0},
}

\usepackage{soul}

\def\be#1\ee{\begin{align}#1\end{align}}

\renewcommand{\dd}{\text{d}}

\newcommand{\p}{\partial}

\renewcommand{\geq}{\geqslant}

\newcommand{\ee}[1]{\end{equation}}
\makeatletter
\let\cat@comma@active\@empty
\makeatother

\begin{document}

%\title{Probing quantum gravity effects at the core of compact objects:the ringdown of semiclassical stars}
%\title{Echoes of quantum gravity: the ringdown of semiclassical stars}
\title{Whispers from the quantum core: the ringdown of semiclassical stars}

\author{Julio Arrechea}
\author{Stefano Liberati}
\author{Vania Vellucci} 
\affiliation{SISSA, Via Bonomea 265, 34136 Trieste, Italy and INFN Sezione di Trieste}
\affiliation{IFPU - Institute for Fundamental Physics of the Universe, Via Beirut 2, 34014 Trieste, Italy}

%----------------------------------------------------------

\begin{abstract}
{This investigation delves into the ringdown signals produced by semiclassical stars, which are ultra-compact, regular solutions of the Einstein equations incorporating stress-energy contributions from quantum vacuum polarization. These stars exhibit an approximately Schwarzschild exterior and an interior composed of a constant-density classical fluid and a cloud of vacuum polarization. By adjusting their compactness and density, we can alter the internal structure of these stars without modifying the exterior. This adaptability enables us to examine the sensitivity of the ringdown signal to the innermost regions of the emitting object and to compare it with similar geometries that differ substantially only at the core. Our results indicate that echo signals are intrinsically linked to the presence of stable light rings and can be very sensitive to the internal structure of the emitting object. This point was previously overlooked, either due to the imposition of reflective boundary conditions at the stellar surface or due to the assumption of low curvature interior geometries. Specifically, for stellar-sized semiclassical stars, we find that the interior travel time is sufficiently prolonged to render the echoes effectively unobservable. These findings underscore the potential efficacy of ultra-compact objects as black hole mimickers and emphasize that any phenomenological constraints on such objects necessitate a detailed understanding of their specific properties and core structure.}
\end{abstract}

%----------------------------------------------------------
%---------------------------------------------
\maketitle
\section{Introduction}
One fundamental difference between the Newtonian and Einsteinian theories of gravity is the presence of upper bounds for the compactness of spherical distributions of matter in hydrostatic equilibrium~\cite{Wald2010GR}. 
While, in Newton's theory, arbitrarily compact constant-density spheres are allowed~\cite{Michell1784,Laplace1789} (by compactness, we refer to the quotient between the total mass and radius of the object), general relativity (GR) shows that the internal pressures of an object are a source of gravity themselves. When these pressures are large enough, they start favoring collapse instead of preventing it, establishing an upper bound for the compactness known as the Buchdahl limit ~\cite{Buchdahl1959}. This limit, in combination with the stability bounds known for neutron stars~\cite{OppenheimerVolkoff1939,RhoadesRuffini1974,FriedmanIpser1987}, and other arguments~\cite{Lynden-BellRees1971, Schmidt1963, Bolton1972}, serve as theoretical guidance suggesting GR black holes as the main candidates for explaining the presently observed massive and ultra-compact objects~\cite{Ghezetal2003, LIGOScientific2016, LIGOScientific2017,EventHorizonTelescope2019, EventHorizonTelescope2022}. Here, by ultra-compact, we mean endowed with a photon circular orbit. 

Nonetheless, current observations still leave room for alternative explanations. In particular, in recent years a considerable attention was devoted to the possibility of distinguishing GR black holes from the alternative regular, singularity free ones predicted by several quantum gravity models~\cite{CarballoRubioetal2018, 1968qtr..conf...87B,Hayward1993,SimpsonVisser2018,Franzin:2023slm}. Another possibility is that the observed objects might be horizonless ultra-compact stars. This idea has been also extensively explored in recent years, leading to the inception of the idea of exotic ultra-compact objects. Of course, were these objects of the stellar kind, the material of which they are composed would necessarily violate the hypotheses behind the Buchdahl theorem. Hence, some sort of new physics beyond classical GR is expected to be necessary for these objects to exists. Among the potentially alternative models, one can cite gravastars~\cite{MazurMottola2004}, fuzzballs~\cite{Mathur2005,Ikedaetal2021}, frozen stars~\cite{BrusteinMedved2017}, and others~\cite{CardosoPani2017, Raposoetal2018, HoldomRen2017}. 

The gravitational signal coming from the coalescence of these compact objects presents potentially detectable features. Indeed, they can exhibit a richer multipolar structure that would modify the inspiral signal at second post-Newtonian order \cite{Pani:2010em,Kesden:2004qx}, as well as non-zero Love numbers that enter the inspiral phase at fifth post-Newtonian order \cite{Cardoso:2017cfl,Herdeiro:2020kba}. Furthermore, being horizonless, these objects present a potentially reflecting surface and an additional stable light ring (see e.g.~\cite{DiFilippo:2024mnc} and references therein), thus they are characterized by a different quasinormal mode (QNM) spectrum and by the presence of echoes in the late-time ringdown signal \cite{Cardoso:2016rao,Vellucci:2022hpl,Franzin:2023slm}.

Preliminary evidence suggesting the presence of GWs echoes in the postmerger stage of binary coalescences observed by Advanced LIGO and Advanced Virgo during the first two observing runs has been reported \cite{Abedi:2016hgu,Conklin:2017lwb,Abedi:2018npz}. Nonetheless, the statistical significance of these GWs echoes has been considered low and compatible with noise \cite{Westerweck:2017hus,Nielsen:2018lkf}. Recently, several additional attempts to analyze the current data %with more sophisticated statistical methods,
still led to no detection \cite{Lo:2018sep,Uchikata:2019frs,Tsang:2019zra}. Additionally, the LIGO/Virgo Collaboration conducted a morphology-independent search for echoes within the events listed in the second and third GWs transient catalogs, but found no statistically significant evidence for echoes (the measurements were consistent with the absence of echoes at the $90\%$ confidence level) \cite{LIGOScientific:2020tif,LIGOScientific:2021sio}. Third-generation gravitational detectors \cite{Punturo:2010zz,Branchesi:2023mws,Kalogera:2021bya,Maggiore:2019uih,Reitze:2019iox,Sathyaprakash:2012jk,LISA:2017pwj,LISA:2022kgy,Barausse:2020rsu} are expected to present higher signal-to-noise-ratio in the ringdown (typically of order O(100)-O(1000)), potentially allowing to detect GWs echoes or to put strong constraints on horizonless models.

In this work we investigate some (often) overlooked aspects related to the ringdown and echoed signal of exotic compact objects. In particular, we study the sensitivity of the signal to the properties of their innermost regions. We present results obtained both in frequency and time domain, highlighting the connection between long living QNMs associated to the presence of a stable lightring --- which in turn is also linked to possible presence non-linear instabilities~\cite{Herdeiro:2020kba,Vellucci:2022hpl} --- and the echoes signal. 

For this investigation we center the analysis on a particular family of static and spherically symmetric exotic compact objects recently found within the framework of quantum field theory in curved spacetimes~\cite{Arrecheaetal2022,Arrecheaetal2023}. These so-called ``semiclassical stars" are a well-motivated model of horizonless compact objects, as they do not require of any new physics beyond quantum vacuum polarization. Such solutions are particular suitable for this analysis since they exhibit an (approximately) Schwarzschild exterior together with a modified interior composed of a constant-density classical fluid and the cloud of vacuum polarization generated by the star itself.
By varying their classical density,  it is possible to modify the interior metric without affecting the exterior, which has a clear effect in the QNM frequencies and the echo waveform. Furthermore, as the compactness of the star is increased, quantum vacuum effects allow to smoothly surpass the Buchdahl limit, opening a window towards studying the properties of exotic compact objects in the black hole limit. 

More specifically, the line element of semiclassical stars has the form
\begin{equation}
    ds^{2}=-f(r)dt^2+h(r)dr^2+r^2d\Omega^2,
    \label{Eq:metric}
\end{equation}
where $d\Omega^{2}$ is the line element of the unit sphere and we define the compactness function as \mbox{$C(r)\equiv2m(r)/r=1-h(r)^{-1}$}, where $m(r)$ is the so-called Misner-Sharp mass~\cite{MisnerSharp1964,HernandezMisner1966}.
In order to find a regular solution, the authors in~\cite{Arrecheaetal2022,Arrecheaetal2023} incorporated the renormalized stress-energy tensor (RSET) of massless scalar fields describing quantum vacuum polarization --- an effect expected to be present in any compact star~\cite{Hiscock1988,ReyesTomaselli2023,Reyes2023} --- to the Einstein equations, so that
\begin{equation}\label{Eq:Semiclassical}
G^{\mu}_{~\nu}=8\pi\left(T^{\mu}_{~\nu}+M_{\rm P}^2\langle\hat{T}^{\mu}_{~\nu}\rangle\right)\,,
\end{equation}
where here and therein we have used geometric units \mbox{$c=G=1$} and thus $\hbar = M_{\rm P}^2$.

Eq.~\eqref{Eq:Semiclassical} takes into account that the physical vacuum of a static star corresponds to the Boulware vacuum~\cite{Boulware1974} due to  quantum field theory (whereas the classical vacuum stress energy tensor would be simply zero). Note that the classical SET of the star matter, which we take to be a constant-density fluid~\cite{Schwarzschild1916b}, and the RSET are conserved independently; they only influence each other through their impact on the spacetime.

Noticeably, by taking equations~\eqref{Eq:Semiclassical}  as a modified theory of gravity to find their self-consistent solutions, the authors~\cite{Arrecheaetal2022,Arrecheaetal2023} obtained a new type of star supported by an interior nucleus of negative energy density. 
When their compactness is small, vacuum polarization contributes as a perturbative correction over the classical constant-density solution. However, as compactness approaches the Buchdahl limit $C_{R}=2M/R=8/9$ (here $M$ is the ADM mass and $R$ the total radius of the star), which the Schwarzschild star saturates~\cite{Buchdahl1959}, the RSET grows in magnitude and modifies the interior structure. As a consequence, semiclassical stars were found to not exhibit an upper bound to their compactness, but rather to be able to exist also for the compactness range $8/9<C_{R}<1$. Thus they can serve as a well-motivated black hole mimicker worth bringing under phenomenological scrutiny.

A distinctive observational feature that only emerges in ultra-compact horizonless objects is the presence of gravitational-wave echoes. Stars with $C_{R}>2/3$ develop outer and inner light rings, which correspond to the presence of unstable and stable circular photon orbits, respectively~\cite{Cardosoetal2014,Cunhaetal2017,DiFilippo:2024mnc}. 
When the system is perturbed and the light-crossing time between the two light rings is large enough, echoes can be individually resolved in the late-time waveform associated to matter fields. In the frequency domain, the signal can also be studied via a complementary analysis involving the QNMs of the system. The time delay between echoes and the associated QNMs are both sensitive to the interior properties of the star~\cite{Cardosoetal2014,Guoetal2022} and, as we will show, a clear correspondence between both these features can be established by analyzing the Discrete Fourier Transform (DFT) of the signal in time domain.
Thus, echo detection~\cite{AbediAfshordi2016} could constitute not only a direct observation of horizonless objects more compact than neutron stars, but also a way to extract details about their interior physical properties. 

Two additional comments are pertinent at this stage. The first comment is related to the presence of an additional scale, $M_{\rm P}$, in the semiclassical equations~\eqref{Eq:Semiclassical}. The introduction of quantum effects breaks the scale invariance of the Schwarzschild interior solution, meaning that the time delay between echoes (and the quasi-normal mode frequencies themselves) does not increase linearly with $M$ as the scale separation between $M$ and $M_{\rm P}$ increases. 
This has clear implications on the detectability of echoes, which, as we will show, would become extremely delayed for semiclassical stars surpassing the Buchdahl limit.

The second comment is related to $r=0$, where absence of curvature singularities implies the metric functions must obey the expansions
\begin{align}\label{Eq:RegMetric}
    h=1+h_{2}r^2+\order{r^3}, \quad  f=f_{0}+f_{2}r^2+\order{r^3}.
\end{align}
We will say that the core of the star is ``de-Sitter-like" (dSl) if $h_{2}<0$ and $f_{2}>0$ and ``anti-de-Sitter-like" (AdSl) if $h_{2}>0$ and $f_{2}<0$ (the cores are only strict de-Sitter or anti-de-Sitter only if $f=h^{-1}$), or ``mixed" otherwise. The characteristics of the central core are relevant for echo detection since, in stars with \mbox{$f_{0}\ll f(R)=1-2M/R$}, the dominant contribution to echo delay comes from a neighbourhood of $r=0$. This property was observed in constant-density solutions~\cite{Zimmermanetal2023} and is preserved in their semiclassical version. Consequently, modifications of the metric in the central regions of the star (for example, changing from an AdSl to a dSl core) have phenomenological consequences. Since the central regions of extremely compact objects would be an adequate place where to search for new physics beyond semiclassical gravity, it is reasonable to think of gravitational-wave echoes as messengers carrying information about the quantum-gravitational regime.

In this article, we present a primer on the phenomenology of semiclassical stars. Although our analysis is clearly model-dependent, the wide range of compactness values and the diversity of interior structures displayed by the semiclassical star model (allowing for AdSl and mixed cores) allows to address more general physical questions about the phenomenology of exotic compact objects. In Sec.~\ref{Sec:SemiStars} we present the classical, uniform-density solution together with its semiclassical counterpart. In Sec.~\ref{Sec:Echoes} we analyze the propagation of spin $2$ test fields perturbations, showing the qualitative differences that arise as the star is made more compact. 
Complementarily, we obtain the fundamental {frequencies of the} quasinormal modes (QNMs) of these stars and their scaling with the compactness in Sec.~\ref{Sec:QNMs}. 
In Sec.~\ref{Sec:SuperCrit} we detail how modifications in the central regions of these stars affects both the delay between echoes and the QNM frequencies. We furthermore discuss about the strong connection between the presence of echoes in the time domain and of long-living modes in the frequency domain and we comment on how both phenomena suggest the necessity of taking into account non-linear effects. We conclude with some further comments in Sec.~\ref{Sec:Conclusions}. 

\section{The semiclassical star model} \label{Sec:SemiStars}
The existence of upper compactness bounds in stars, as well as their specific values, depends on the properties and equation of state obeyed by the fluid~\cite{Karageorgis2007,Andreasson2008,UrbanoVeermae2018,Raposoetal2019}. The simplest stellar model, or constant-density star~\cite{Schwarzschild1916b}, saturates the hypotheses of the Buchdahl theorem and its compactness is bounded from above by $C_{R}<8/9$. For clarity, we briefly review the model here. 

The stress-energy tensor (SET) of the star is modeled as an isotropic perfect fluid
\begin{equation}\label{Eq:ClassicalSET}   T^{\mu}_{~\nu}=\left(\rho+p\right)u^{\mu}u_{\nu}+p\delta^{\mu}_{\nu},
\end{equation}
with $p$ and $\rho$ denoting the pressures and energy density measured by an observer comoving with the fluid with $4$-velocity $u^{\mu}$. Assuming a metric of the form \eqref{Eq:metric} the covariant conservation of the SET~\eqref{Eq:ClassicalSET} implies
\begin{equation}\label{Eq:Cont}\nabla_{\mu}T^{\mu}_{~r}=p'+\frac{f'}{2f}\left(\rho+p\right)=0.
\end{equation}
If we assume a uniform energy density fluid
\begin{equation}\label{Eq:EoS}
    \rho(r)\equiv\rho=\text{const},
\end{equation}
we can solve Eqs.~\eqref{Eq:Semiclassical} with $M_{\rm P}=0$ and seek for  a metric  matching the Schwarzschild one at the star surface $r=R$. Such interior metric can be found to be
\begin{align}\label{Eq:SchwStar}
    ds^{2}=
    &
    -\frac{1}{4}\left(3\sqrt{1-C_{R}}-\sqrt{1-r^{2}C_{R}/R^{2}}\right)^{2}dt^{2}\nonumber\\
    &
    +
    \left(1-r^{2}C_{R}/R^{2}\right)^{-1}dr^{2}
    +r^{2}d\Omega^{2}.
\end{align}
It is straightforward to check from the above equation that $f=0$ at $r=0$ for $C_{R}=8/9$. By means of the conservation relation~\eqref{Eq:Cont} which, for constant-density fluid, can be integrated to
\begin{equation}
    p=\rho\left(\sqrt{\frac{f(R)}{f(r)}}-1\right),
\end{equation}
this translates into a diverging pressure at the center, which is tantamount to say that for this solution, any finite pressure will not be able to support the star beyond the $C_{R}=8/9$ compactness. This is, in a nutshell, the so called Buchdahl's limit.

In what follows, while still assuming the equation of state~\eqref{Eq:EoS} for the classical fluid, we shall be interested in solving the full semiclassical Eqs.~\eqref{Eq:Semiclassical} incorporating the RSET as an additional matter source. Obtaining the RSET in spherically symmetric spacetimes is a complicated procedure requiring, beyond renormalization, also an accurate numerical computation for the modes in which the quantum field is decomposed. While efforts have been devoted to calculating RSETs in black hole spacetimes~\cite{CandelasHoward1984,Andersonetal1995,LeviOri2016,Taylor:2022sly}, this computation is yet to be attained in stellar spacetimes if not via analytical approximations. While exact RSETs can be found numerically, this still hinders a proper treatment of their backreaction on the classical geometry~\cite{FlanaganWald1996}.

Following this motivation, in~\cite{Arrecheaetal2023} a novel approximation --- particularly adapted to stellar spacetimes --- for the RSET of massless minimally coupled scalar fields  was developed. Remarkably, within such framework semiclassically stable star were found. Here, we omit the technical details, as they can be easily found in the original publication, and just present the form of the $tt$ and $rr$ components of the so obtained semiclassical stellar
equations~\eqref{Eq:Semiclassical}
\begin{widetext}
    \begin{align}\label{Eq:SemiTT}
        \frac{h'}{h}=
        &
        -\frac{h-1}{r}+8\pi r h \rho+\frac{M_{\rm P}^2}{48\pi r}\left\{-\frac{83}{20 r^2 h}+\frac{1}{15 r^2}\left(105+136\pi r^2\rho+504\pi r^2 p\right)\right.\nonumber\\
        &
        \left.-\frac{h}{30 r^2}\left[63+1054\pi r^2\rho+7488\pi^2 r^4\rho^2+49\pi r^2\left(53+940\pi r^2 \rho\right)p+56832\pi^2 r^4 p^2\right]
        \right.\nonumber\\
        &
        \left.-\frac{h^2}{15 r^2}\left(1+8\pi r^2 p\right)^2\left(3-616\pi r^2\rho-1080\pi r^2 p\right)-\frac{11h^3}{20r^2}\left(1+8\pi r^2 p\right)^4\right.\nonumber\\
        &
        \left.+8\pi \left(\rho+p\right)\log\left[\lambda^2 f\right]\left[-1+h\left(1+2\pi r^2 \rho+26\pi r^2 p\right)- h^2\left(1+8\pi r^2 p\right)^2\right]\right\},\\
        \frac{f'}{f}=
        &
        \frac{h-1}{r}+8\pi r h p+\frac{{M_{\rm P}^2}}{48\pi r}\left\{
        -\frac{5}{4r^2 h}+\frac{1}{15r^2}\left(21-108\pi r^2 \rho-764\pi r^2 p\right)\right.
        \nonumber\\
        &
        \left.+\frac{h}{30r^2}\left[23-4\pi r^2\rho\left(3+20\pi r^2 \rho\right)+320\pi r^2\left(5-12\pi r^2 \rho\right)+11904\pi^2 r^4 p^2\right]\right.\nonumber\\
        &
        \left.-\frac{h^2}{15 r^2}\left(1+8\pi r^2 p\right)^2\left(11-180\pi r^2\rho-228\pi r^2 p\right)
        -\frac{11h^3}{60 r^2}\left(1+8\pi r^2 p\right)^4\right.\nonumber\\
        &
        \left.+4\pi \left(\rho+p\right)\log\left[\lambda^2 f\right]\left[\frac{11}{3}-\frac{h}{3}\left(3+12\pi r^2\rho+20\pi r^2 p\right)-h^2\left(1+8\pi r^2 p\right)^2\right]
        \right\},\label{Eq:SemiRR}
    \end{align}
\end{widetext}
where $\lambda$ is an arbitrary parameter, introduced by the renormalization prescription, which was implicitly defined as
\begin{equation}\label{Eq:NuValue}
    \log \left[\lambda^{2}f(R)\right] = \frac{h(R)\left[15h(R)-40\pi R^2\rho-6\right]-9}{h(R)\left[5h(R)+24\pi R^2\rho+6\right]-11}
\end{equation}
so to allow for a smooth matching between the interior and exterior solutions~\cite{Arrecheaetal2023}.

A quick glance at the above equations shows that the right-hand sides of Eqs.~(\ref{Eq:SemiTT},~\ref{Eq:SemiRR}) contain corrections proportional to $M_{\rm P}^2$ that qualitatively estimate quantum vacuum polarization effects. For stars of small compactness, for which $p\ll \rho$, semiclassical effects amount to a perturbative correction over the classical solution. However, when compactness approaches the Buchdahl limit, in which $p\gg \rho$, terms $\propto p^{4}$ in~(\ref{Eq:SemiTT},~\ref{Eq:SemiRR}) overcome their $\order{M_{\rm P}^2}$ suppression, modifying the stellar solutions in a non-perturbative way and allowing to surpass the Buchdahl limit.

Eqs.~(\ref{Eq:SemiTT},~\ref{Eq:SemiRR}) reduce to their counterparts in vacuum by taking $p=\rho=0$. At sufficiently large distances, the metric obeys the expansions
\begin{align}\label{Eq:AsymptExp}
    f
    &
    =1-\frac{2M}{r}+\frac{M_{\rm P}^2 M^2}{90\pi r^4}+\order{\frac{M_{\rm P}^2 M^3}{r^5}},\nonumber\\
    h
    &
    =\left[1-\frac{2M}{r}-\frac{M_{\rm P}^2 M^2}{6\pi r^4}+\order{\frac{M_{\rm P}^2 M^3}{r^5}}\right]^{-1}.
\end{align}
Notice that the semiclassical solution is asymptotically flat because we are evaluating the RSET in the Boulware state, which reduces to the Minkowski vacuum in the asymptotic regions.
With these boundary conditions, the vacuum equations are integrated (assuming $M>0$) from a sufficiently large fiducial radius until some radius $R>2M$ where the surface of the star is placed. Then, starting from $r=R$, Eqs.~(\ref{Eq:SemiTT},~\ref{Eq:SemiRR}) are integrated inwards with the boundary conditions
\begin{align}
    f=f(R),\quad h=h(R),\quad \left. p\right|_{r=R}=0,\quad \rho=\mbox{const}=\rho_{0},
\end{align}
where the value of $\rho_{0}$ compatible with a regular metric must be found numerically~\cite{Arrecheaetal2023}. The surface compactness of the star is given by $C_{R}=1-h(R)^{-1}$ and can take values in the interval $C_{R}\in(0,1)$. 

Not all values of $\rho_{0}$ are compatible with a regular metric: numerical explorations reveal that for \textit{any} $C_{R}<1$ there exists a critical density $\rho_{0}=\rho_{\text{c}}$ that corresponds to the regular solution with smallest $\rho_{0}$. For $\rho_{0}<\rho_{\text{c}}$ we find solutions with naked curvature singularities, {whereas for $\rho_{0}\geq\rho_{\text{c}}$ we obtain fully regular spacetimes. Since the space of parameters is large, we have summarized these cases in Table~\ref{Table}.} In the following sections, our analyses will be restricted to critical stars $(\rho_{0}=\rho_{\text{c}})$, which display a core of the AdSl type. The phenomenology associated to super-critical stars $(\rho_{0}>\rho_{\text{c}})$, which have cores of the mixed type instead, will be explored in Section~\ref{Sec:SuperCrit}.
\begin{table}[htb]
\centering\includegraphics[width=0.6\textwidth]{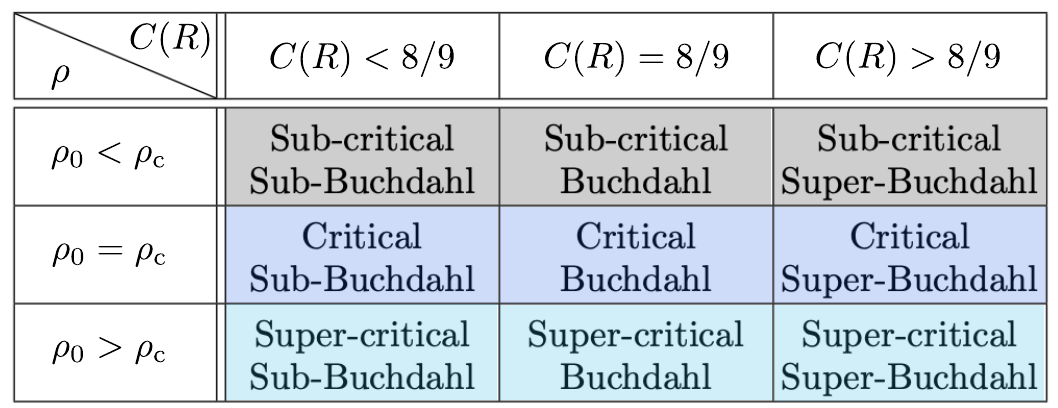}\label{Table}\caption{Table depicting the various families of semiclassical stars, depending on their energy density $\rho_{0}$ and their surface compactness $C_{R}$. The grey cells correspond to singular solutions. In dark blue (critical solutions), we have regular stars with an AdSl core, while in light blue (super-critical solutions) we find stars with mixed cores. In both cases the solutions can approach the black hole limit arbitrarily.}
\end{table}

\begin{figure}[htb]
    \centering
    \includegraphics{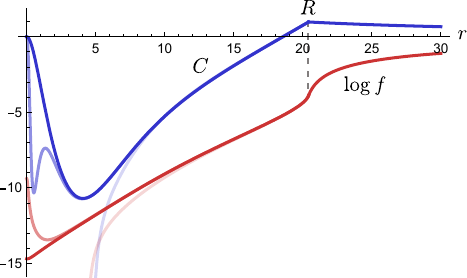}
    \caption{Numerical solution of semiclassical stars with $C_{R}=0.98,~M/M_{\rm P}=10$. Their surface is located at $R\approx 20.42 M_{\rm P}$ and the spacetime for $r>R$ closely resembles the Schwarzschild vacuum solution. The blue curve is the compactness function $C=1-h(r)^{-1}$ and the red curve corresponds to $\log{f(r)}$. From darker to lighter shades: critical solution with $\rho=\rho_{\text{c}}\simeq0.00112 M_{\rm P}^{-2}$, sub-critical solution with $\rho/\rho_{\text{c}}\simeq 0.99$ and super-critical solution with $\rho/\rho_{\text{c}}\simeq 1.00015$.
    The distinctive features of semiclassical stars are their negative mass interiors (recall that $C=2m(r)/r$) and their monotonously decreasing $f(r)$, which produces large redshifts on outgoing null rays. Critical and super-critical solutions have cores of the AdSl and mixed types, respectively.}
    \label{Fig:Metric}
\end{figure}

Figure~\ref{Fig:Metric} shows the metric functions of a critical semiclassical star with $M/M_{\rm P}=10$ and $C_{R}=0.98$, far surpassing the Buchdahl limit. 
Note that, in this work, we take stars of such small sizes to  simplify numerical analyses. Nonetheless, we checked that the interior properties of semiclassical stars do not suffer any qualitative modification for larger $M/M_{\rm P}$ ratios. Therefore, we can safely extrapolate the conclusions of this study to stars of astrophysical size.

As compactness is increased beyond \mbox{$C_{R}=8/9$}, the mass function inside the stars solutions becomes increasingly negative. This effect is a direct consequence of the negative energy densities generated by the RSET, which grow in negativity as central pressures increase. All in all, semiclassical stars are a well-motivated extension of constant-density stars beyond their maximum compactness bound. Once vacuum polarization effects allow the Buchdahl threshold to be surpassed, we find no further compactness limits. In the upcoming sections we will explore the phenomenology associated to the propagation of test fields on these stars.
\begin{figure}
    \centering
    \includegraphics[width=0.5\textwidth]{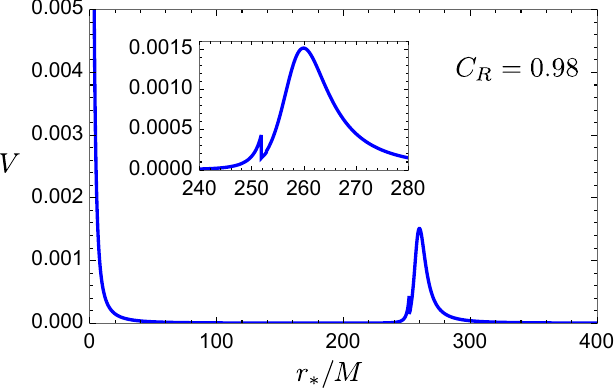}
    \caption{Potential of the test-field equation for $s=l=2$ and compactness $C_R=0.98$. The inset is a zoom on the discontinuity of the potential at the star's surface. The finite jump in the potential indicates the position of the star surface. Such discontinuity is linked with the finite jump in $\rho$ at the surface, and poses no problem for the numerical evolution of test fields.}
    \label{fig:potential}
\end{figure}

\section{Critical solutions: Time domain analysis}
\label{Sec:Echoes}
In this section we study the time evolution of a test field in the spacetime described in Section \ref{Sec:SemiStars}. {Since we are using a test-field approximation, we are neglecting any coupling between our perturbation field and the perturbations of the matter present in our spacetime. For uniform density stars in GR, the axial matter perturbations vanish~\cite{ThorneCampolattaro1967,ChandrasekharFerrari1991, Nollert1999}, this means that test-field perturbations with $s=2$ describe accurately the axial sector of gravitational waves. Since for compactness $C_{R}<8/9$ semiclassical corrections are perturbative, we expect the test field limit to be a good approximation also in the semiclassical case.
For stars with $C_{R}>8/9$ instead, the RSET is non-negligible and the coupling between its perturbations and our field perturbation could lead to non-negligible corrections. Unfortunately, it is not possible at the moment to investigate this further since our RSET approximation corresponds to a vacuum-expectation value obtained under the assumptions of staticity and spherical symmetry, and is thus not adequate in time-dependent or non-spherical situations. However, we expect that our qualitative conclusions about test-field perturbations will be valid also for the true coupled gravitational fields.}

Because of the spherical symmetry, we can decompose the field into the radial part and angular part, the latter being expressed in terms of spherical harmonics. Then the equation for the radial part of scalar, electromagnetic and gravitational test-field perturbations is ~\cite{Karlovini:2001fc,Medved:2003pr}:
\be
\frac{\p^2 \psi}{\p t^2} -\frac{\dd^2 \psi_s}{\dd r_*^2} +  V_s \psi_s = 0\,,
\label{radialEq}
\ee
\begin{comment}
Because of the spherical symmetry, the Klein–Gordon equation is separable and thus we can decompose the field as $\Phi(t,r,\theta,\varphi)=\sum_{lm}\frac{\Psi_{lm}(t,r) }{r} Y_{lm}(\theta,\varphi)$, where $Y_{lm}$ are the scalar spherical harmonics.
Then the field equation for each mode $\Psi_{lm}(t,r)$ is (to avoid cluttering, in what follows we drop the $lm$ indexes):
\be
\frac{\p^2 \Psi}{\p t^2} - \frac{\p^2 \Psi}{\p r_*^2} +V(r) \Psi=0\,,\label{radialKG}
\ee
\end{comment}
where the tortoise coordinate $r_*$ is defined as
\begin{equation}
    dr_*=\sqrt{\frac{h}{f}} dr,
\end{equation}
and $s$ is the spin of the test-field, $l$ is the harmonic index and the potential reads
\be
V(r)=\frac{f(r)s(s-1)}{r^2 h(r)}+\frac{f(r)(l(l+1)-s(s-1))}{r^2}
+\frac{(s-1)(h'(r)f(r)- h(r)f'(r)}{2 r h(r)^2}\label{Eq:Potential}
\ee

The shape of $V(r)$ is reported in Fig.~\ref{fig:potential} for a reference value of the compactness.
We solved Eq. (\ref{radialEq}) using a fourth-order Runge–Kutta integrator and computing spatial derivatives
with finite-difference approximation of second-order in accuracy.
In the numerical simulation reported in this section we always consider an $l=2$ quadrupolar mode and we use as initial condition for $\psi$ a Gaussian pulse:
\be
\psi(r,0)= \psi_0 \exp\left(-\frac{(r_*-r_*^c)^2}{2\sigma^2}\right), \quad \\ \frac{\p \psi(r,0)}{\p t}=-\psi_0 \frac{(r_*-r_*^c)}{\sigma^2} \exp\left(-\frac{(r_*-r_*^c)^2}{2\sigma^2}\right),
\ee
%(or a gaussian wave-packet $\Psi(r,t)= a e^{-k(r_*-r_c+t)^2 /(2 c^2)+i \omega (r_*+t)}$ ),
with central value $r_*^c=r_*(r^c)=r_*(250 M)$ and width $\sigma=2M$;  different initial values lead to similar results.
The pulse is initially centered outside the peak of the potential $V(r)$, and moves inwards.

\subsection{Time delay between echoes}
For sufficiently compact semiclassical stars, as for every other ultra-compact object with an exterior region well-approximated by (or exactly described by) the Schwarzschild metric, the signal in time domain is given by an initial response pulse that is very similar (or identical) to the one expected from Schwarschild black holes followed by a series of smaller packets usually called echoes~\cite{Cardoso:2017cqb, Annulli:2021ccn}. The distance between these echoes is given by the light-crossing time between the potential peak and the reflective boundary. In some simplified models, the perturbation is assumed to be reflected at the surface of the object, leading to a logarithmic dependence of the light-crossing time (aka time-delay) with $C_{R}^{-1}-1$ ~\cite{CardosoPani2017}. 
However, a complete reflection of gravitational waves at the surface is not realistic and one expects instead that the perturbation (or a part of it) can travel through the interior of the object. In spherical symmetry, this translates into a reflective boundary condition at the center $r=0$.

In this case the full formula for the time delay between echoes (assuming a Schwarzschild exterior) reads \cite{CarballoRubioetal2018} 
\begin{equation}
    T_{\rm echo}=2M {-4M(C_R^{-1}-1)}- 4M\ln{\left[2\left({C_R^{-1}-1}\right)\right]}+T_{\rm int}\,,
\end{equation}
where the first {three} terms on the r.h.s.~are associated to the travelling time between the surface and the light ring while $T_{\rm int}$ is the travelling time to cross the star. Normally such time has been neglected in the extant literature assuming a low curvature interior. However this is not the case for our semiclassical solutions where indeed in the interior a very large (but finite) time delay is suffered by exiting waves. This can be easily seen by a direct calculation using the modified geometry.

The light-crossing time a light ray emanating from the photon sphere and getting reflected at $r=0$ needs to reach the photon sphere back is
\begin{equation}
\tau_{\text{ph}}=2\int_{0}^{r_{\text{ph}}}\sqrt{\frac{h}{f}}dr
\end{equation}
When $f(0)\ll f(R)$ and $f$ decreases monotonically towards the center, the dominant contribution to the crossing time comes from the large time delays suffered by light rays at the innermost regions of the star. This implies that the logarithmic dependence of the crossing time on $C_{R}^{-1}-1$ (that came exclusively from the exterior of the object) becomes a sub-leading contribution.

The magnitude of the time delay for the semiclassical star model depends on both the compactness $C_{R}$ and the ratio $M/M_{\rm P}$ between the mass of the object and the Planck mass $M_{\rm P}$, which enters the solution through the renormalized stress-energy tensor. Semiclassical corrections introduce a new length scale that spoils the scale invariance of the classical solution.

\begin{figure}[htb]
\centering\includegraphics[width=0.5\textwidth]{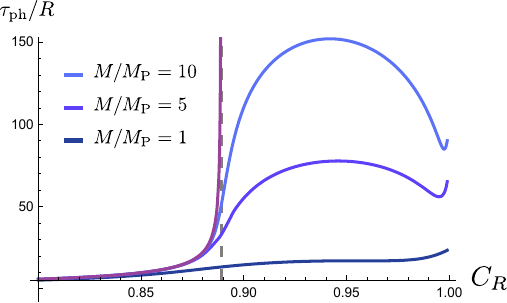}  \caption{Crossing time in terms of the compactness for the classical constant density star (purple line) and semiclassical stars with $M/M_{\rm P}=\left\{10,5,1\right\}$ (blue lines from top to bottom, respectively). Semiclassical solutions are not scale invariant. The origin of the plateau reached by the light crossing time for sufficiently high compactness is discussed in the text. }
    \label{Fig:CrossingTime}
\end{figure}

\begin{figure}[H]
\begin{center}
\includegraphics[scale=0.9]{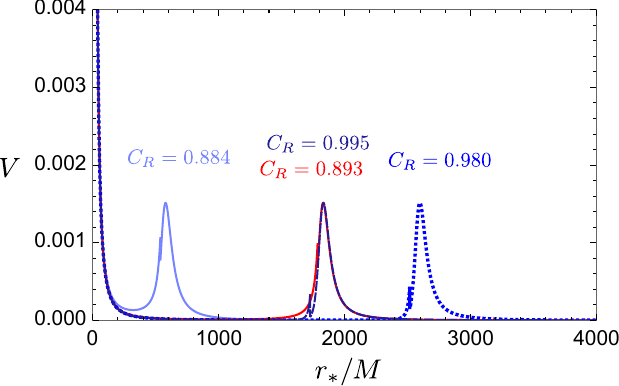}
\caption{ Potential of the test-field equation for a star with $M/M_P=10$ and different compactness. We can see that for sufficiently high compactness the distance between the peak and the central barrier at $r_*=0$ start to decrease leading to a smaller crossing time (a smaller time delays between echoes). For this reason, it is possible to have the same crossing time for different values of the compactness (e.g. for $C_R=0.995$ and $C_R=0.893$). However, also in these cases, the potentials remain slightly different and thus the spectrum of QNMs for the two compactness, albeit very similar, will not be perfectly degenerate. % Right panel: potential of the test-field equation of a star with $M/M_P=10$ but different compactness and classical density value (with critical density in dashed line and super-critical density in continuous line). The two configurations present the same distance between the peak and the central barrier  at $r_*=0$ and thus the same time-delay between echoes, however they present a different slope of the potential in the interior region. As we will see in section \ref{Sec:SuperCrit}, this brings to differences in the echos signal. 
}
\label{fig:Potentials}
\end{center}
\end{figure}
Figure~\ref{Fig:CrossingTime} displays the qualitative dependency of the crossing time on both compactness and scale separation. Regarding the dependency on the compactness, we observe an initial growth of the crossing time with $C_{R}$ until a maximum value is reached, followed by a subsequent decrease. 
This behavior is caused by the decrease of values of the functions $h$ and $f$ in the stellar interior as $C_{R}$ increases. We can associate decreases in $f$ with larger time delays, while decreases in $h$
can be linked to a shrinking of the proper volume. These effects are perceived by null rays and compensate themselves causing, beyond certain $C_{R}>8/9$ value, a decrease of the crossing time with the compactness. Representations of the potentials both for sub-Buchdahl and super-Buchdahl stars are shown in Fig.~\ref{fig:Potentials}. It is possible to generate two stars of different compactness that have the same crossing time. The shape of their associated potentials will be different nonetheless, leading to distinct observational signatures. 

Regarding the dependency on scale separation between the ADM mass and the Planck mass, for the realistic case of a  stellar mass object, $M/M_{\rm P}\gg1$, we see that for low values of the compactness, $C_{R}<8/9$, the quantity $\tau_{\text{ph}}/R$ approaches the classical solution value and is hence basically independent on $M/M_{\rm P}$. For very compact objects, $C_{R}> 8/9$, the crossing time $\tau_{\text{ph}}/R$  grows instead linearly with $M/M_{\rm P}$.  In between both regimes lies the separatrix $C_{R}=8/9$, for which $\tau_{\text{ph}}/R\propto\left(M/M_{\rm P}\right)^\alpha$, with $\alpha\in(0,1)$. These scalings are valid for all the three cases of sub-critical, critical and super-critical density. This scaling of the crossing time, as we will see below, has crucial implications for the observability of the echo signal. 

%When $M\gg M_{\rm P}$, for sub-Buchdahl stars we recover the scaling $\tau_{\text{ph}}/R\propto \text{const.}$, whereas for super-Buchdahl stars we find $\tau_{\text{ph}}/R\propto M/M_{\rm P}$. The Buchdahl solution separates both regimes, showing a scaling of the form \mbox{$\tau_{\text{ph}}/R\propto\left(M/M_{\rm P}\right)^\alpha$}, with $\alpha\in(0,1)$. This scaling of the crossing time, as we will see below, has crucial implications for the observability of the echo signal. 

\begin{figure}[h!]
\includegraphics[scale=0.3]{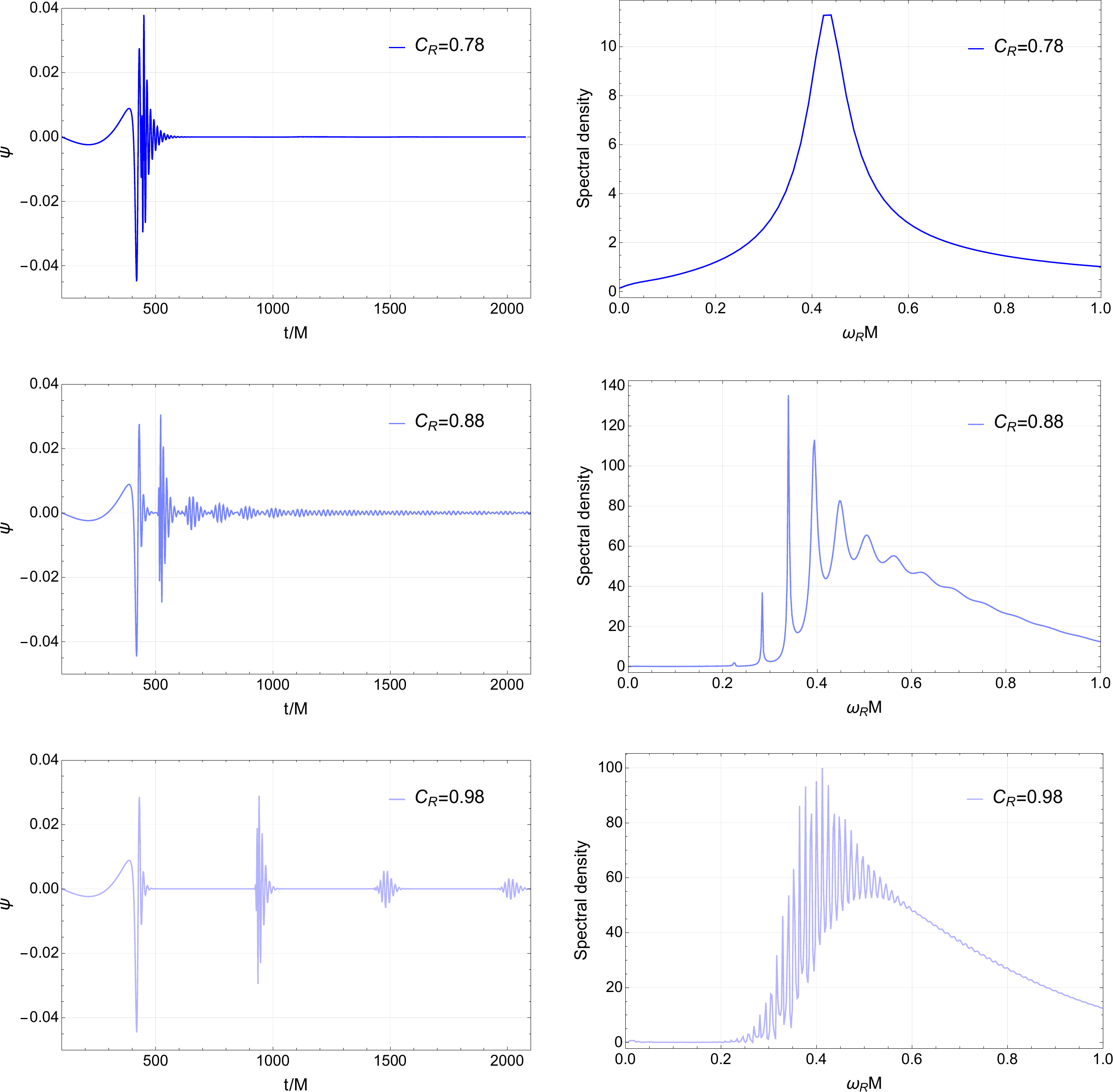}
    \caption{Signal received by an observer at $r=250 M$ outside a semiclassical star with $M/M_{\rm P}=10$ for different values of the compactness $C_R$. The initial signal is a Gaussian pulse centered outside the potential peak. We observe an initial Schwarzschild-like signal associated to the peak of the potential followed by a series of echoes, whose separation increases with the compactness. We show the obtained time domain signal in the left panels and the corresponding Discrete Fourier transform (DFT) of the echoes part of the signal (from $t/M=500$) in the right panels. Note that for smaller compactness a lot of modes are not resolved in the DFT because they decay too quickly and so they present a low spectral amplitude (compared to that of the longest living modes) in the time interval we are analyzing. As instance, for the $C_R=0.78$ we can resolve only the fundamental (longest living) mode.}
    \label{fig:echoes1}
\end{figure}

Figure~\ref{fig:echoes1} shows three numerical integrations of Eq.~\eqref{radialEq} for stars with the same $M/M_{\rm P}$ and increasing compactness, all of them exhibiting an outer light ring and leading, therefore, to an initial black hole-like response followed by a sequence of echoes. As it is customary for exotic compact objects, 
time delay between echoes grows with the compactness (although this is not always true, see Fig.~\ref{Fig:CrossingTime} and Fig. \ref{fig:Potentials}), and they become clearly identifiable as isolated events when the star is super-Buchdahl. After the initial black-hole like response, the subsequent echoes would suffer such a large time delay that would make them essentially impossible to identify as part of the original signal~\cite{Zimmermanetal2023}, at least for super-Buchdahl objects with masses much larger than the Planck mass.  These signals are contrasted with the Discrete Fourier Transforms (DFT) obtained by Fourier transforming the portion of the signal containing echoes. The peaks of the DFTs correspond to the individual frequencies of the QNMs. As the compactness increases, so does the size of the cavity in which the field is propagating, and the DFT allows to individually resolve more trapped modes.

%, but in this model, when $C_{R}>8/9$, the crossing time grows approximately quadratically with $M/M_{\rm P}$.
%This has crucial implications for the observability of the echo signal. After the initial black-hole like response, the subsequent echoes would suffer such a large time delay that would make them essentially impossible to identify as part of the original signal~\cite{Zimmermanetal2023}, at least for objects with masses much larger that the Planck mass. 
%For example, if the compactness is $\approx 0.90$, values of $\tau_{ph}$ of order of seconds that lead to detectable echoes are possible only for values of M of order of $10^{21} \sqrt{h}$ that is $10^{-17}$ solar masses. We conclude then that the only possible way to observe echoes from this model of semiclassical stars is to have partial reflection at the surface or to observe the signal coming from extremely light primordial objects. 

Of course the observed scaling of the crossing time with size is a feature of this particular model of semiclassical stars, which is nonetheless shared by other semiclassically-inspired objects~\cite{Chen:2024ibc}.
Already within our semiclassical set-up it is possible to construct different configurations starting from classical matter fluids obeying other, more realistic equations of state instead of using the simple constant density profile. It is not clear yet if this particular scaling of the time delay found here will be shared by semiclassical stars with different equations of state for the classical sector.

%\begin{figure}[h!]
% \includegraphics[scale=0.25]{DifferentM.pdf}
%    \caption{Signal received by an observer at $r=250 M$ outside a semiclassical star with compactness $C_R=0.90$ for different values of the mass M in terms of the Planck length. The initial signal is a gaussian pulse centered outside the potential peak. }
%    \label{fig:echoesM}
%\end{figure}
\begin{comment}

\begin{figure}
 \includegraphics[scale=0.3]{SemiclassCR0.99.pdf}
    \caption{Signal received by an observer at $r_*=110 M$ for a given value of the compactness $C_R=0.99$. In the upper panel the perturbation is completely reflected at the surface of the star while in the bottom panel it travels through the entire star until it is reflected by the centrifugal barrier at the center. The initial signal is a gaussian pulse centered outside the potential peak. TO REDO STARTING WITH A GAUSSIAN AT THE SAME R}
    \label{fig:reflect}
\end{figure}
\end{comment}
\begin{comment}

\begin{figure} 
\includegraphics[scale=0.2]{SemiclassCR0.99A0.90.pdf}
    \caption{Signal received by an observer at $r_*=110 M$ for a given value of the compactness $C_R=0.99$. The initial signal is a gaussian pulse centered outside the potential peak. Once arrived at the surface part of the perturbation is reflected back to infinity while another part (the 90 $\%$) continues travelling through the star until it is scattered back to infinity by the centrifugal barrier at the center. TO REDO  STARTING WITH A GAUSSIAN AT THE SAME R}
    \label{fig:partreflect}
\end{figure}
\end{comment}

\section{Critical solutions: Frequency domain analysis}\label{Sec:QNMs}
\begin{comment}
Quasinormal modes dictate the behaviour of perturbations propagating on given spacetimes. For $s=2$ fields, such perturbations belong either to the axial or polar sector, depending on whether they impart a differential rotation to the system or not. Both types of perturbations can be analyzed as the scattering of scalar fields by a potential barrier. For simplicity, we are going to focus in axial perturbations in the test field limit, that is, ignoring any excitation of the fluid modes. 
This corresponds to considering only the so-called spacetime modes~\cite{ChandrasekharFerrari1991,Anderssonetal1995,Anderssonetal1996}. For uniform density stars in GR, the fluid perturbations vanish in the specific case of axial perturbations~\cite{ThorneCampolattaro1967,ChandrasekharFerrari1991, Nollert1999}. Since for compactness $C_{R}<8/9$ semiclassical corrections are perturbative, we expect the test field limit to be a good approximation also in the semiclassical case. However, for stars with $C_{R}>8/9$, the RSET is non-negligible and its perturbations would need to be incorporated. Unfortunately, it is not clear how to do this since our RSET approximation corresponds to a vacuum-expectation value obtained under the assumptions of staticity and spherical symmetry, and is thus not adequate in time-dependent or non-spherical situations.
Under these assumptions, axial quasinormal modes correspond to the solutions of the field equation~\eqref{radialEq} for $s=2$ fields assuming an ansatz of the form 
\end{comment}

{We shall now move to study the same $s=2$ test-field perturbations in the frequency domain. This means that we will solve Eq.~\eqref{radialEq} assuming an ansatz of the form }
\begin{equation}
    \psi(t,r)=e^{i\omega t}\phi(r),
\end{equation}
where $\phi$ obeys the boundary conditions
\begin{align}
    \phi
    &
    \simeq e^{i\omega r},\quad r\to\infty,\nonumber\\
    \phi
    &
    \simeq r^{1+l},\quad r\to0.
\end{align}
Quasi-normal mode frequencies correspond to the frequency values $\omega=\omega_{\text{R}}+i\omega_{\text{I}}$
for which the solution $\phi$ satisfies both boundary conditions simultaneously. 

Numerous methods have been developed for obtaining the quasinormal modes of black hole and stellar spacetimes (see~\cite{KokkotasSchmidt1999,Konoplyaetal2019} and references therein), whose utility depends on the details of the system under consideration.
Here, we implement the direct calculation developed by Chandrasekhar and Detweiler~\cite{ChandrasekharDetweiler1975}. While its implementation is straightforward, this method suffers from instabilities that originate from a sensitivity in the solution to the radius where the ingoing boundary conditions are imposed~\cite{Molinaetal2010}. Despite its limitations, it is possible to generate reliable and accurate results for the axial QNMs of semiclassical stars and to compare them with their classical counterparts. Below we give further details on the way we tested the precision of our results. Throughout this section we take $M/M_{\rm P} = 1$. 

For what regards the boundary conditions at large $r$, let us remember that our metric is characterized by the asymptotic behaviour set by the expansions~\eqref{Eq:AsymptExp}. Henceforth, for the ingoing mode we have assumed an asymptotic series solution of the form
\begin{equation}
    \phi=e^{i\omega r}\left(\frac{r}{2M}-1\right)^{2M i \omega}\sum_{n=0}^\infty \frac{c_{n}}{r^n},
\end{equation}
and solved for the $c_{n}$ coefficients order by order, so obtaining
\begin{align}
    c_{1}
    &
    =\frac{l\left(1+l\right)i}{2\omega}c_{0},\nonumber\\
    c_{2}
    &
    =\frac{\left(1-l\right)l\left(1+l\right)\left(2+l\right)+12 M i\omega}{8\omega^{2}}c_{0},
\end{align}
as the lowest-order coefficients, with $c_{0}=1$.

At small $r$, the metric obeys instead the expansions~\eqref{Eq:RegMetric} and we can then find the following series solution for the outgoing mode
\begin{equation}
\phi=r^{l}\sum_{m=1}^{\infty}d_{m}r^{m},
\end{equation}
where, for the lowest-order terms, we obtain
\begin{align}
    d_{2}
    =0,\quad d_{3}
    &
    =-\frac{\omega^{2}+\left(f_{2}-f_{0}h_{2}l\right)\left(2+l\right)}{2f_{0}\left(3+2l\right)}d_{1},\nonumber\\
    d_{4}
    &
    =-\frac{-3f_{3}+f_{0}h_{3}\left(1+2l\right)}{12f_{0}}d_{1},
\end{align}
with $d_{1}=1$.

\begin{figure}
\centering\includegraphics[width=0.5\textwidth]{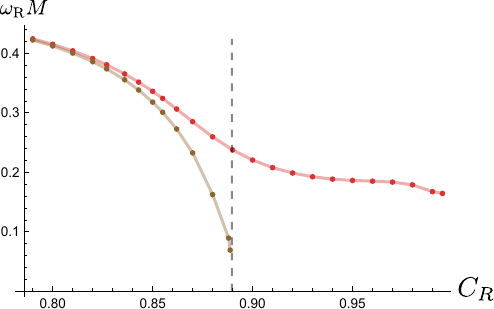}
    \caption{Real part of the QNM frequencies in terms of the compactness for the classical and semiclassical models (brown and red, respectively). The classical frequencies approach $0$ in the Buchdahl limit, while the semiclassical ones have finite values in all the range $C_{R}<1$. We have taken $M=M_{\rm P}=1$.}
    \label{Fig:ReQNM}
\end{figure}

\begin{figure}
\centering\includegraphics[width=0.5\textwidth]{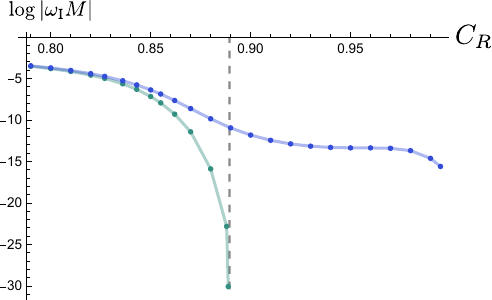}
    \caption{Logarithm of the imaginary part of the QNM frequencies in terms of the compactness for the classical and semiclassical models (green and blue, respectively). The classical frequencies approach $0$ in the Buchdahl limit, while the semiclassical ones have finite values in all the range $C_{R}<1$. The presence of small imaginary part is indicative of long-lived modes and potential non-linear instabilities~\cite{Franzin:2023slm}. Notice the similarity in behaviour with respect to the real parts in Fig.~\ref{Fig:ReQNM}. We have taken $M=M_{\rm P}=1$. }
    \label{Fig:ImQNM}
\end{figure}

Using these boundary conditions, we shift the value of $\omega$ until we find the one that guarantees a vanishing of the Wronskian between the ingoing and outgoing modes at some radius $R_{\text{m}}>R$. We calculated the modes with a numerical precision such that the Wronskian vanished at $R_{\text{m}}$ up to 12 decimal places. For every mode computed, we checked that this vanishing was satisfied for different matching radii $R_{\text{m}}$, thus verifying that the frequencies obtained were numerically stable. In this way, we obtained the fundamental frequency of the $l=2$ mode and observed how its value changes with the compactness. We want to stress that for fundamental frequency we mean the one with the smallest imaginary part of all the star spectrum and not the one closest to the black hole fundamental mode. Figures~\ref{Fig:ReQNM} and~\ref{Fig:ImQNM} show the real part and the logarithm of the imaginary part of the fundamental (longest living) QNM frequency, respectively. The corresponding values of the classical, constant density solution are shown for comparison. We have tested our method comparing it with the results in~\cite{ChandrasekharFerrari1991} for constant density stars, with which we find good numerical agreement. For classical stars, we observe a drop in the accuracy of the method as we approach the Buchdahl limit, precisely where the small imaginary part of the QNM becomes comparable to the value of the Wronskian, but the QNM frequencies of the last two points shown in Figs.~\ref{Fig:ReQNM} and~\ref{Fig:ImQNM} are nonetheless accurate within one order of magnitude. For semiclassical stars, the QNM frequency values obtained are accurate at least to 3 decimal places under large variations in the matching radius $R_{\text{m}}$ in the entire range of compactness values explored. 

It is easy to see that the QNM frequencies of the classical and semiclassical solutions are characterized by quite distinct behaviours. In the first case, both $\omega_{R}$ and $\omega_{I}$ tend to $0$ in the Buchdahl limit. This behaviour is associated to the divergence of the crossing time in this limit, which produces an infinite time delay between echoes~\cite{Zimmermanetal2023}. In the semiclassical solution, however, the QNM frequencies start departing from their classical counterparts as the Buchdahl limit is approached, since in proximity of such limit semiclassical effects become relevant. As discussed, the latter also lead to a breakdown of scale invariance introducing a dependence of the frequencies on $M/M_{\rm P}$. 
%Hence, we expect the values of $\omega_{R}M$, and $\omega_{I}M$ to depart from their classical values for $C_{R}\gtrsim 8/9$. 
Finally, let us notice that between the Buchdahl and the black hole limits, axial QNM frequencies reach a plateau, not increasing significantly. The maximum compactness value we considered in this study was $C_{R}=0.995$. In the $C_{R}\to1$ limit, these solutions develop a curvature singularity at their surface (a feature that strongly depends on the RSET approximation~\cite{Arrecheaetal2022b}), thus we avoided taking values too close to the black hole compactness here.

\begin{figure}
    \centering
    \includegraphics[width=1\linewidth]{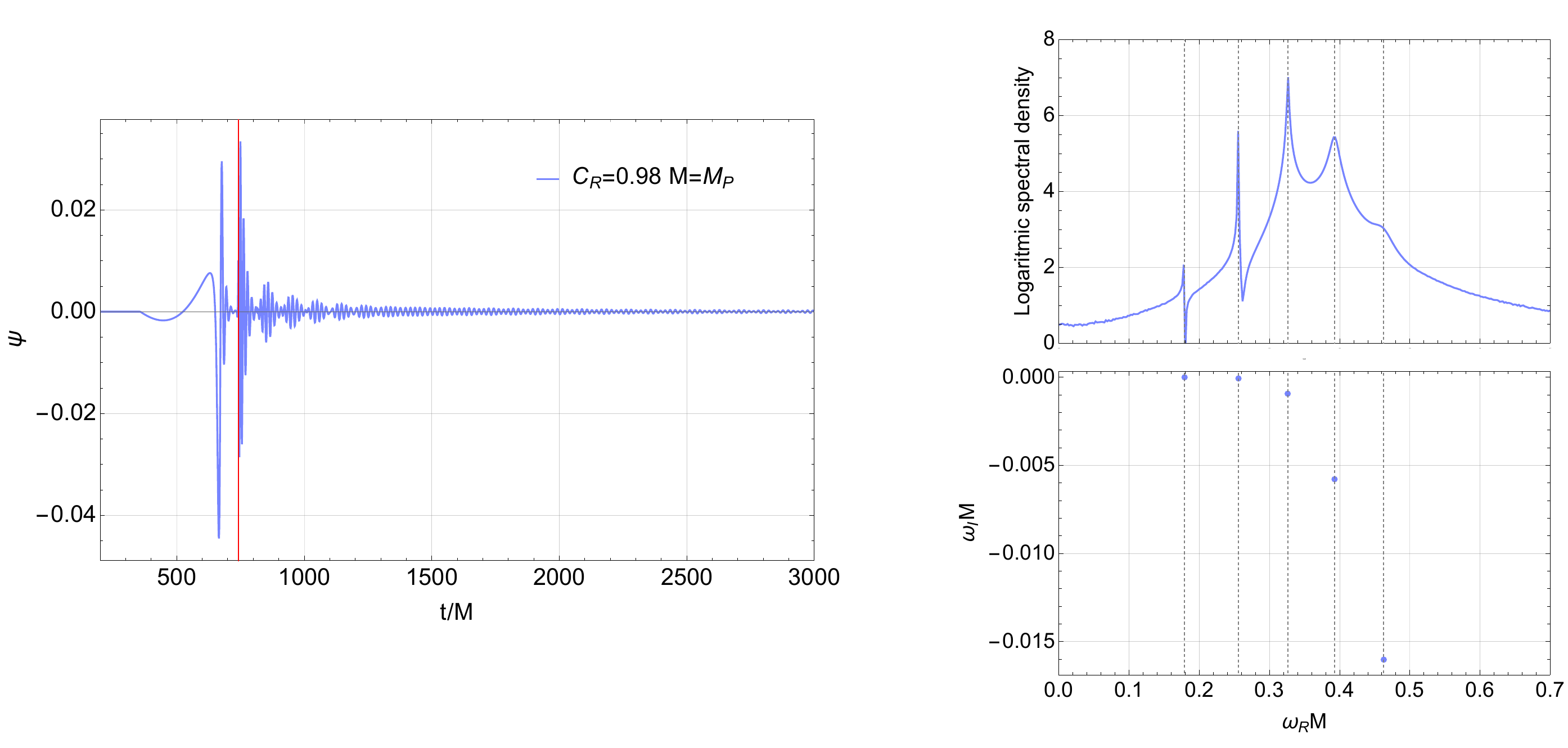}
    \caption{Comparison between the trapped modes obtained from the frequency domain analysis (right bottom panel) and the ones obtained through the DFT (right top panel) of the time domain signal (left panel). The DFT is performed only on the echoes part of the time-domain signal (the part after the red line in the left panel plot) in order to clearly show the longest living (trapped) modes. }
    \label{fig:confronto}
\end{figure}

The modes obtained with this frequency analysis are in agreement with the ones we find through a DFT of the echoes signal obtained in time domain, as it is showed for the  specific example in figure \ref{fig:confronto}. 
%The classical solutions are scale invariant, which causes $\tau_{R}/R$, $\omega_{R}M$, and $\omega_{I}M$ to stay constant under changes in $M$ for fixed $C_{R}$. In the semiclassical case, this scale invariance is broken due to the introduction of another scale $\hbar$, resulting in $\tau_{R}/R$, $\omega_{R}M$, and $\omega_{I}M$ to vary with $M/M_{\rm P}$. Note that for a larger separation of scales, we expect the semiclassical frequencies to stay closer to the classical ones, making the mean frequency values in the $8/9<C_{R}<1$ region smaller. 

\section{Sensitivity to the internal properties of compact objects}
\label{Sec:SuperCrit}
Phenomenological studies of black hole mimickers spacetimes attempt to be agnostic about the interior properties of such objects by imposing a (partial or total) reflective boundary condition at their surface~\cite{CardosoPani2019,Maggioetal2021, Maggio2023}. This restricts echoes and, in general, the ringdown signal, to be only probes of the spacetime geometry between the surface and the light ring. However, it is clear that a more physical scenario would allow for perturbations to travel through the entire object, and consequently to carry out information about its internal structure. For a spherically symmetric spacetime this assumption would translate into reflective boundary conditions being imposed at the center of the object rather than at its surface. Remarkably, by doing so, the ringdown signal becomes also a probe of the innermost regions where quantum effects are expected to be stronger. 

In this section we compare the phenomenology of compact objects with the same exterior spacetime but different central regions, showing that echoes and QNMs are indeed sensitive to the internal structure and hence indirectly to the quantum effects avoiding singular behaviours at the object core. As particular examples, we take a sub-family of semiclassical stars with increasing classical energy density and the Dymnikova metric.

\subsection{Super-critical semiclassical stars}
For a star of a given mass $M$ and compactness $C_{R}$, the semiclassical stellar equations~(\ref{Eq:SemiTT},~\ref{Eq:SemiRR}) allow to freely adjust the value of the classical density $\rho_0$. In previous sections, all our considerations have been restricted to critical solutions, i.e., those with $\rho_0=\rho_{\text{c}}$ (see Table~\ref{Table}), for which the central pressure is a global maximum and the region near $r=0$ is of the AdSl type. By taking $\rho_0>\rho_{\text{c}}$, we can generate regular stellar solutions of the same mass and compactness, but whose central pressure has a local minimum at the center, this feature being reproduced by cores of the dSl type. %{\bf Can we show this behaviour of the pressure in a graph?}

\begin{figure}[hbt]
    \centering
    \includegraphics[width=0.4\textwidth]{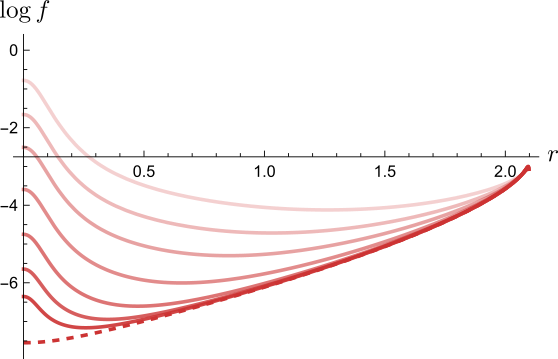}
    \caption{Logarithm of the redshift function $f$ in terms of the radius for stars with $M=M_{\rm P}=1$ and $C_{R}=0.99$ . The dashed line denotes the critical solution with $\rho_{\text{c}}\simeq0.055 M_{\rm P}^{-2}$. From darker to lighter shades, the curves correspond to super-critical solutions with $\rho_0/\rho_{\text{c}}=\left\{1.001,1.004,1.015,1.04,1.1,1.25,1.6\right\}$. Increasing $\rho_0$ diminishes the redshift suffered by light rays at the center appreciably.}
    \label{Fig:SuperCritF}
\end{figure}

\begin{figure}[htb]
    \centering
    \includegraphics[width=0.4\textwidth]{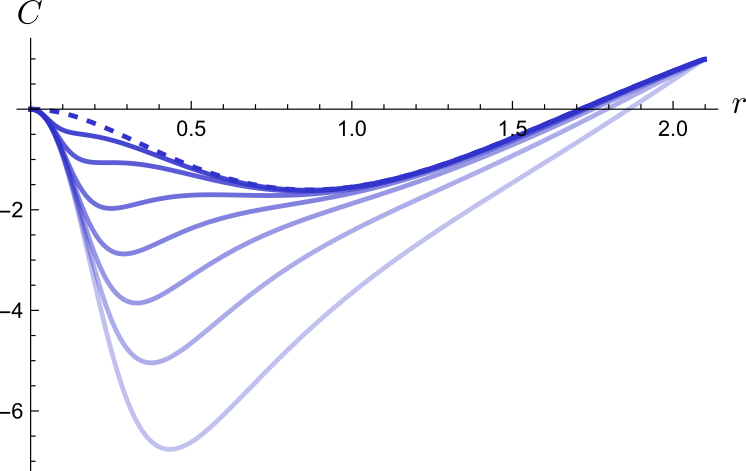}
    \caption{Compactness function $C$ in terms of the radius for stars with $M=M_{\rm P}=1$ and $C_{R}=0.99$. The dashed line denotes the critical solution with $\rho_{\text{c}}\simeq0.055 M_{\rm P}^{-2}$. From darker to lighter shades, the curves correspond to super-critical solutions with $\rho_0/\rho_{\text{c}}=\left\{1.001,1.004,1.015,1.04,1.1,1.25,1.6\right\}$. Increasing $\rho_0$ generates a deeper and broader negative mass interior.}
    \label{Fig:SuperCritC}
\end{figure}

Figures~\ref{Fig:SuperCritF} and~\ref{Fig:SuperCritC} show respectively {the interior metric function $f(r)$ and $C(r)=1-h(r)^{-1}$} for several super-critical solutions. Note that all super-critical solutions have an identical vacuum exterior and only differ in their interior properties.
\begin{figure}[hbt]
    \centering
\includegraphics[width=0.5\linewidth]{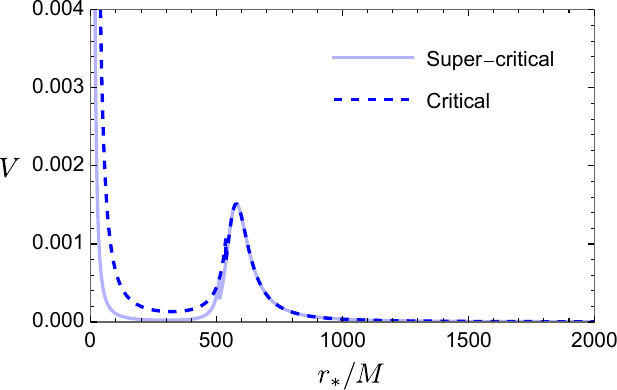}
    \caption{Comparison between the potential of the test-field equation for stars with $M/M_P=10$, same light-crossing time but different density parameters. The dashed line correspond to the potential for the critical value of the density while the continuous one is for the super-critical density $\rho=1.026679 \rho_{c}$. The two potentials have a different trend in the internal region of the object between the lighting peak and the reflective barrier at $r=0$.}
    \label{fig:PotenSuper}
\end{figure}

\begin{figure} [htb]
    \centering    
    \includegraphics[width=0.4\textwidth]{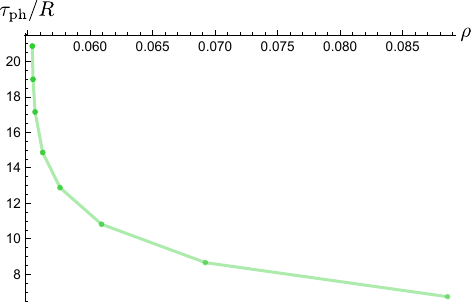}
    \caption{Light-crossing time in terms of the density of super-critical solutions with $M=M_{\rm P}=1$ and $C_{R}=0.99$. As the star becomes super-critical, the crossing time decreases appreciably. This behaviour affects the QNM frequencies, increasing their values with $\rho$ (see Figs.~\ref{Fig:ReFreqSuperCrit} and ~\ref{Fig:ImFreqSuperCrit}).}
    \label{Fig:CrossingTimeSuperCrit}
\end{figure}

\begin{figure}[htb]
    \centering
    \includegraphics[width=0.4\textwidth]{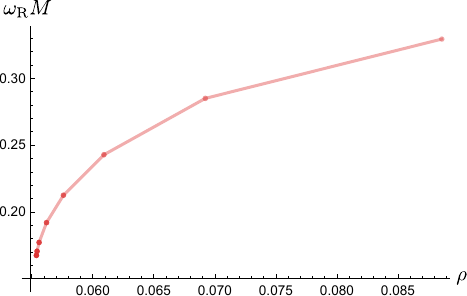}
    \caption{Real part of the $l=2$ fundamental QNM frequencies for super-critical solutions with $M=M_{\rm P}=1$ and $C_{R}=0.99$. The frequency values increase with $\rho$, similarly to the logarithm of the complex part of the QNM frequency and the inverse of the crossing time (see Figs.~\ref{Fig:ImFreqSuperCrit} and ~\ref{Fig:CrossingTimeSuperCrit})}
    \label{Fig:ReFreqSuperCrit}
\end{figure}

\begin{figure}[htb]
    \centering    
    \includegraphics[width=0.4\textwidth]{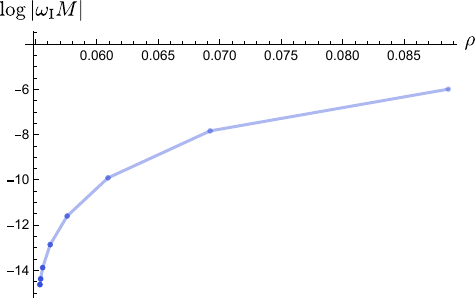}
    \caption{Logarithm of the complex part of the $l=2$ fundamental QNM frequencies for super-critical solutions with $M=M_P=1$ and $C_{R}=0.99$. The frequency values increase with $\rho$ in a tendency very similar to the real part of the frequency and the inverse of the crossing time (see Figs.~\ref{Fig:ReFreqSuperCrit} and ~\ref{Fig:CrossingTimeSuperCrit}).}
    \label{Fig:ImFreqSuperCrit}
\end{figure}

Although there is no change in the position of the outer light ring, the modification of the stellar interior affects the crossing time (hence the time delay between echoes) as well as the QNM frequencies. Particularly, the crossing time diminishes as $\rho_0$ increases (Fig.~\ref{Fig:CrossingTimeSuperCrit}), while the QNM frequencies increase (Figs.~\ref{Fig:ReFreqSuperCrit} and~\ref{Fig:ImFreqSuperCrit}). 

Indeed, inserting expansions~\eqref{Eq:RegMetric} in the potential~\eqref{Eq:Potential} for $s=2,~l=2$, we have
\begin{equation}\label{Eq:Potentialr0}
    V=\frac{6f_{0}}{r^2}+\order{r}^0,
\end{equation}
hence different densities correspond to different values of $f_{0}$, meaning changes in the slopes of the centrifugal barrier at $r=0$ (see figure \ref{fig:PotenSuper}) that, as shown in Fig.~\ref{Fig:InternalComparison}, affect the phase of the echoes and cause a slightly different modulation of their amplitude with time.

\subsection{Dymnikova's model}
The Dymnikova metric was originally proposed as a regular black hole candidate~\cite{Dymnikova:1992ux,Dymnikova:2001fb}. However, it was recently realized that, for any regular black hole metric, a corresponding horizonless counterpart can be obtained for a sufficiently large value of the singularity-regularization scale $\ell$. While we could assume the latter to be of the order of the Planck mass, it is not constrained to be so, and actually quite larger values have been considered in the extant literature. Note however that, for this model, the range of possible scale separations between M and $\ell$ is limited. Indeed, as anticipated, if $\ell/M$ is too small the object presents a pair of horizons (as it is indeed a regular black hole), and if $\ell/M$ is too big the object has no light rings. For this reason the range of parameters for which echoes appear is restricted.

More specifically, for such model the line element has the same form of Eq.~\ref{Eq:metric} with
\begin{equation}
    h(r)=1/f(r)\,,\qquad f(r)=1-\frac{2 m(r)}{r}\,,\qquad \mbox{and} \quad m(r)=M \left[1 - \exp(-\frac{r^3}{2M\ell^2})\right]\,.
\end{equation}
%With $\ell$ a generic mass scale responsible for the regularization of the central singularity.

The exterior solution of the Dymnikova's model closely resembles the one of our semiclassical model. Both models have nearly the same outer light ring and innermost stable circular orbit positions, and the potential for test-field perturbations shares the same shape from the neighborhood of the light ring to infinity. However, the interior region is quite different. 
Near the center, Dymnikova's metric exhibits a strictly de-Sitter core and the potential for test-field perturbations has a different shape (w.r.t.~the semiclassical star)  especially near $r=0$ (see Eq.~\eqref{Eq:Potentialr0}). 

\begin{figure}[htb]
    \centering   
     \includegraphics[scale=0.30]{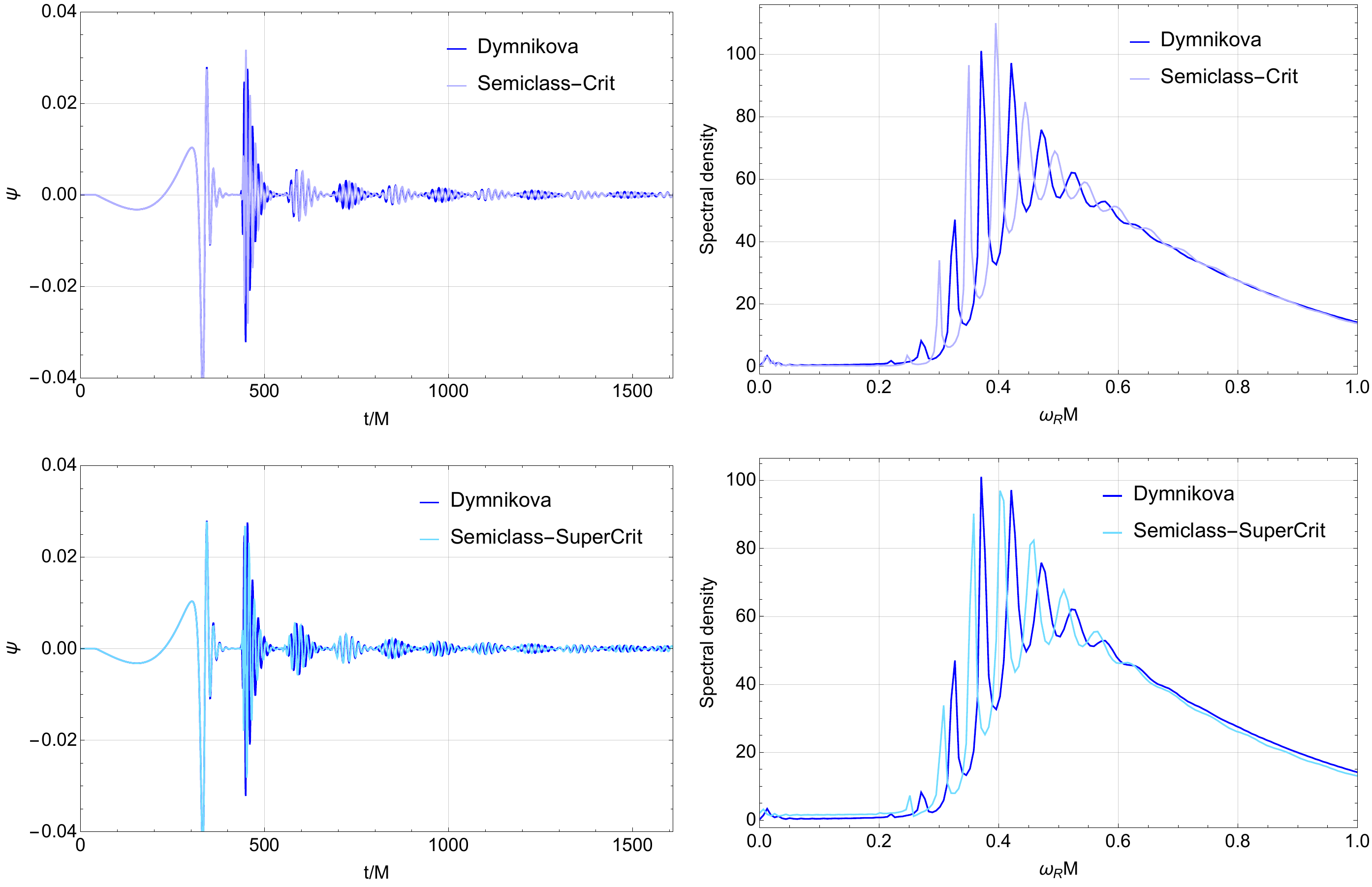}
    \caption{Comparison between signals received by an observer at $r=250 M$ outside objects with similar exterior spacetime but different central region and $M/M_{\rm P}=10$. The initial signal is a gaussian pulse centered outside the potential peak. The Dymnikova model with $\ell=1.15 M$ 
    %before was 1.38 M
    is compared with semiclassical stars with critical density $\rho_{c}$ in the upper panels and with super-critical density $\rho=1.026679 \rho_{c}$ in the bottom panels. In the time domain plots (right panels) we can note, in both cases, a slight difference in the modulation of the amplitude and a different phase between the two signals. These differences can be better appreciated in the frequencies domain, indeed the FFT of the echoes part of the signals (right panel) show that the compared configurations present different trapped modes. This is due to a slightly different trend of the potential in the interior of the object, especially in the region near $r=0$ (see figure \ref{fig:PotenSuper} as instance).  }
    \label{Fig:InternalComparison}
\end{figure}

As a further tests of the sensitivity of echos to the internal structure of the ultra-compact star we can compare the Dymnikova model with semiclassical star solutions characterized by the same light-crossing time between the unstable light ring and the center, i.e. having by construction the same time delay between echoes. Nonetheless, as shown in Fig.~\ref{Fig:InternalComparison}, one can see that the time-domain signal coming from stars with different densities presents some differences within them and with Dymnikova's model. Indeed, signals coming from different models present slightly different modulations in amplitude and very different phases that are due, mainly, to the discrepancies in their potential barrier near $r=0$.
%Even when we compare solutions with the same light-crossing time between the unstable light ring and the center, the signal coming from a compact object described by the Dymnikova metric presents some differences with respect to the signal coming from semiclassical stars belonging both to the critical and super-critical families. Even when we compare solutions with the same light-crossing time between the light ring and the center, meaning that the time delay between echoes is the same, the time-domain signal coming from stars with different densities presents some differences, as shown in Fig.~\ref{Fig:InternalComparison}. We can observe in Figure~\ref{Fig:InternalComparison} that signals coming from different models present slightly different modulations in amplitude very different phases that are due, mainly, to the discrepancies in the potential barrier near $r=0$. %\vv{From Stefano: Say more, but what?}
%\vv{Maybe we can add a quantitative difference in the QNMs of solutions with same exterior but different center for example Critical VS Super Critical or VS Dym. In other words, is it a detectable difference at least theoretically?}

We therefore conclude that black hole mimickers spacetimes with similar exterior geometry, i.e.~nearly sharing the same position for the external light ring, but differing in their interior properties will still produce different ringdown signals. Indeed, this realization can be seen as the ultra-compact star analogue of the even more striking one that regularizations of black hole cores lead to solutions which generically have  (even if possibly just slightly) different exteriors w.r.t.~the singular solutions of GR (at least modulo nonphysical ad hoc constructions).

\subsection{Echoes and the stable lightring}\label{lightechoes}

We have observed that the waveform of echoes strongly depends on the internal structure of the object. This is because this part of the signal is associated with the slow leakage of perturbations from the region between the unstable light ring and the central centrifugal barrier (where a stable light ring must always be present~\cite{Cunha:2017qtt,DiFilippo:2024mnc}). This same process leads to the appearance of long-living modes in the QNM spectrum, which are considered indicative of non-linear instabilities~\cite{Cardoso:2014sna,Cunha:2017qtt}. These instabilities appear to be confirmed in some specific models by numerical simulations \cite{Cunha:2022gde}. 

Typically, these concerns are not associated with the presence of echoes in the signal. However, it is worth stressing that echoes are the time-domain counterpart of the aforementioned long-living modes \cite{Tominaga:1999iy,Kokkotas:1995av,Ferrari:2000sr} and can indeed be reconstructed using the first few overtones~\cite{Guo:2022umh}, with the late-time part of the echo signal being dominated by the fundamental modes. Therefore, it is not surprising that a careful analysis of the echo scenario also leads to the conclusion that non-linear effects and backreaction must play an important role \cite{Vellucci:2022hpl,Chen:2019hfg}.

It is important to note that the long-living modes forming the echoes are due to the extended time required for perturbations to travel between the unstable light ring and the reflective barrier. Thus, they can lead to backreaction only in the case of a sustained continuous flux of perturbations, allowing for a, possibly destabilizing, \textit{accumulation} of energy in the region between the unstable light ring and the reflective barrier. There is no physical reason to expect a true \textit{amplification} of perturbations, even at the non-linear level, as it is unclear how the latter could be compatible in this setting with energy conservation (as no ergoregion is present is such horizonless geometries).

Finally, we emphasize that while the extremely long time delay between echoes and their consequently probable unobservability (discussed in section~\ref{Sec:Echoes}) is a model-dependent feature, the presence of a stable light ring and the consequent semi-trapping of perturbations is a universal feature of horizonless objects and thus a necessary condition for the emission of echoes. Therefore, studying the ringdown in a fully non-linear regime would be very useful to understand if we can truly observe echoes and thus probe the innermost regions of these objects.

\section{Conclusions}
\label{Sec:Conclusions}
In this work, we studied the ringdown signal produced by semiclassical stars, which are regular stellar solutions to the Einstein equations incorporating the expectation value in the Boulware vacuum (the natural vacuum in stellar spacetimes) of the renormalized stress-energy tensor. A distinctive feature of these stars is that the backreaction of vacuum polarization effects allows for the existence of regular stars surpassing the Buchdahl limit. Consequently, they can serve as a novel family of black hole mimickers without requiring new physics beyond GR and quantum field theory in curved spacetimes.

The exterior of these objects corresponds to a perturbatively-corrected Schwarzschild spacetime, maintaining the same asymptotically flat behavior at large distances up to some surface $r=R$. However, in the interior region $r<R$, if the star's compactness surpasses the Buchdahl limit, the spacetime is significantly modified by semiclassical corrections.

Notably, by varying their compactness $C_{R}$ and their classical surface density $\rho_0$, we can modify the internal region of these stars without changing the exterior. This means that the position of the outer light ring and the innermost stable circular orbit remain the same as in the Schwarzschild black hole (up to $\order{M_{\rm P}^2}$ corrections). This allowed us to investigate the extent to which the ringdown signal is sensitive to the internal structure of black hole mimickers.

Our time-domain analysis shows that the initial part of the signal is identical to that produced by a Schwarzschild black hole,: information about the object being a black hole mimicker is only carried by the subsequent echoes. Furthermore, we observed that increasing the compactness of the object generally results in echoes being further separated and individually resolved. However, for sufficiently compact objects, the crossing time can decrease with compactness (see Fig.~\ref{Fig:CrossingTime}).

The latest part of the echo signal is dominated by the long-living fundamental QNM frequencies, which are highly sensitive to the features of the interior metric. We varied these features by increasing the compactness of critical solutions (Figs.~\ref{Fig:ReQNM},~\ref{Fig:ImQNM}) and by exploring super-critical solutions with the same $C_{R}$ but different $\rho_0$ (Figs.~\ref{Fig:CrossingTimeSuperCrit},~\ref{Fig:ReFreqSuperCrit},~\ref{Fig:ImFreqSuperCrit}). From our analysis, we conclude that echoes unmistakably carry information about the internal properties of these objects, making even more pressing the question of whether such echoes are truly observable.

We have shown that if perturbations can travel through the object's interior, the time-delay between echoes depends on both the compactness of the object and its internal structure. It also crucially depends on the ratio between the ADM mass $M$ and the Planck mass. For super-Buchdahl stars ($C_{R}>8/9$), the ratio $\tau_{\text{ph}}/R$ increases linearly with the quotient $M/M_{\rm P}$, making echoes essentially unobservable for stellar-sized objects. For example, with $C_{R}\approx 0.90$, values of $\tau_{ph}$ of the order of seconds --- that might lead to detectable echoes --- are possible only for $M\approx 10^{21} M_{\rm P}$, or roughly $10^{-17}$ solar masses. For {a stellar mass} object, the part of the signal that crosses the exterior potential barrier and travels towards the object's interior would, in practice, be frozen forever. 

Nonetheless, possible mechanisms for observable echoes from such semiclassical stars could include partial reflections at their surface~\cite{Press1979,Bemficaetal2020} or ringdown signals from extremely light primordial objects~\cite{Carr:1975qj,Arbey:2024ujg} or morsels formed during binary mergers ~\cite{Cacciapaglia:2024wtp,Chitishvili:2021jrx}.  Also, it might be possible to find other observables sensitive to the internal structure of these semiclassical stars, for example in the inspiral-merger part of the signals. We hope that these issues will be investigated in future work.

It is important to stress that the extremely delayed echo signals here found, could be just a feature of this particular model, where semiclassical effects break the scale invariance of the classical uniform-density solution. Within the same semiclassical framework, other models can be constructed starting from different classical stress-energy sources. E.g.~the approximations used for the RSET can be applied to fluids with barotropic equations of state~\cite{Arrecheaetal2023}. Investigating whether echoes are extremely delayed and thus unobservable in these other models would be a desirable investigation to carry on in the future. 

Another aspect, not considered in this study, is the effect of matter on the signal, as we analyzed test-field perturbations. In the axial sector, classical matter perturbations vanish, but the RSET is non-zero (albeit small) even outside the star's surface, potentially affecting the initial part of the ringdown signal through interactions with quantum matter. Trust in the results obtained with test field perturbations follows from such effects being of $\order{M_{\rm P}^2}$, but a direct check would be valuable.

Regarding possible observations, we would like to stress that it is also possible to fit the quasi-normal modes we obtained (and the time-delay between echoes) as functions of the star parameters so that, from the observed signal, it would be possible to infer some properties of the potential and of the underlying  metric (similiarly to what happens for neutron stars \cite{Andersson:1997rn}). Of course, this requires to assume this specific model among all the possible exotic compact objects and, in addition, this procedure can present some degeneracies. For example the time delay between echoes can be mapped to the length of the cavity between the potential peak and central barrier at $r_*=0$, but as explained in section III.A and V, different choice of the mass, compactness and classical density can lead to the same cavity length. Nonetheless, we do expect that future observations of multiple modes and large numbers of events will be able to break such degenercies.

Finally, despite these various possible future lines of investigation, we want to emphasize a key takeaway from our study of this specific class of semiclassical ultra-compact stars: i.e.~that they provide us with the general lesson that the core of these exotic objects does matter and that once propagation inside the star core is considered, short-delayed (and potentially observable) echoes are no longer guaranteed. We can see this conclusions ``as a curse and a blessing": on the downside, it is evident that testing these objects via echoes will require in the future an in-depth analysis based on the specific exotic star model. On the upside, if such analysis will yield promising results for phenomenology, it will open the door to finally testing quantum gravity effects --- {hidden in the core of such objects} --- by observing gravitational waves. Only time will tell.

%Other classical fluid spheres suffer from similar (more stringent) compactness limits~\cite{Andreasson2008,Alhoetal2022}. 
%It seems likely that the crossing time diverges when such limits are saturated, since an infinite crossing time is the consequence of a diverging pressure at $r=0$ (which denotes the compactness limit itself). In that case, if semiclassical effects allow to generically surpass these compactness limits, it seems likely that it will produce exotic compact objects with extremely large crossing times. 
%Whether vacuum polarization provides a truly generic mechanism for extending hydrostatic equilibrium beyond the limits imposed by GR would be a direction worth pursuing in the future. 
%\newpage
\bibliography{bib-refs}

%apsrev4-2.bst 2019-01-14 (MD) hand-edited version of apsrev4-1.bst
%Control: key (0)
%Control: author (8) initials jnrlst
%Control: editor formatted (1) identically to author
%Control: production of article title (0) allowed
%Control: page (0) single
%Control: year (1) truncated
%Control: production of eprint (0) enabled
\begin{thebibliography}{109}%
\makeatletter
\providecommand \@ifxundefined [1]{%
 \@ifx{#1\undefined}
}%
\providecommand \@ifnum [1]{%
 \ifnum #1\expandafter \@firstoftwo
 \else \expandafter \@secondoftwo
 \fi
}%
\providecommand \@ifx [1]{%
 \ifx #1\expandafter \@firstoftwo
 \else \expandafter \@secondoftwo
 \fi
}%
\providecommand \natexlab [1]{#1}%
\providecommand \enquote  [1]{``#1''}%
\providecommand \bibnamefont  [1]{#1}%
\providecommand \bibfnamefont [1]{#1}%
\providecommand \citenamefont [1]{#1}%
\providecommand \href@noop [0]{\@secondoftwo}%
\providecommand \href [0]{\begingroup \@sanitize@url \@href}%
\providecommand \@href[1]{\@@startlink{#1}\@@href}%
\providecommand \@@href[1]{\endgroup#1\@@endlink}%
\providecommand \@sanitize@url [0]{\catcode `\\12\catcode `\$12\catcode `\&12\catcode `\#12\catcode `\^12\catcode `\_12\catcode `\%12\relax}%
\providecommand \@@startlink[1]{}%
\providecommand \@@endlink[0]{}%
\providecommand \url  [0]{\begingroup\@sanitize@url \@url }%
\providecommand \@url [1]{\endgroup\@href {#1}{\urlprefix }}%
\providecommand \urlprefix  [0]{URL }%
\providecommand \Eprint [0]{\href }%
\providecommand \doibase [0]{https://doi.org/}%
\providecommand \selectlanguage [0]{\@gobble}%
\providecommand \bibinfo  [0]{\@secondoftwo}%
\providecommand \bibfield  [0]{\@secondoftwo}%
\providecommand \translation [1]{[#1]}%
\providecommand \BibitemOpen [0]{}%
\providecommand \bibitemStop [0]{}%
\providecommand \bibitemNoStop [0]{.\EOS\space}%
\providecommand \EOS [0]{\spacefactor3000\relax}%
\providecommand \BibitemShut  [1]{\csname bibitem#1\endcsname}%
\let\auto@bib@innerbib\@empty
%</preamble>
\bibitem [{\citenamefont {Wald}(2010)}]{Wald2010GR}%
  \BibitemOpen
  \bibfield  {author} {\bibinfo {author} {\bibfnamefont {R.}~\bibnamefont {Wald}},\ }\href {https://books.google.it/books?id=9S-hzg6-moYC} {\emph {\bibinfo {title} {General Relativity}}}\ (\bibinfo  {publisher} {University of Chicago Press},\ \bibinfo {year} {2010})\BibitemShut {NoStop}%
\bibitem [{\citenamefont {Michell}(1784)}]{Michell1784}%
  \BibitemOpen
  \bibfield  {author} {\bibinfo {author} {\bibfnamefont {J.}~\bibnamefont {Michell}},\ }\bibfield  {title} {\bibinfo {title} {{On the Means of Discovering the Distance, Magnitude, \&c. of the Fixed Stars, in Consequence of the Diminution of the Velocity of Their Light, in Case Such a Diminution Should be Found to Take Place in any of Them, and Such Other Data Should be Procured from Observations, as Would be Farther Necessary for That Purpose.}},\ }\href {https://doi.org/10.1098/rstl.1784.0008} {\bibfield  {journal} {\bibinfo  {journal} {Phil. Trans. Roy. Soc. Lond.}\ }\textbf {\bibinfo {volume} {74}},\ \bibinfo {pages} {35} (\bibinfo {year} {1784})}\BibitemShut {NoStop}%
\bibitem [{\citenamefont {Laplace}(2009)}]{Laplace1789}%
  \BibitemOpen
  \bibfield  {author} {\bibinfo {author} {\bibfnamefont {P.-S.}\ \bibnamefont {Laplace}},\ }\href {https://doi.org/10.1017/CBO9780511693335} {\emph {\bibinfo {title} {Exposition du systéme du monde}}},\ \bibinfo {edition} {2nd}\ ed.,\ Cambridge Library Collection - Mathematics\ (\bibinfo  {publisher} {Cambridge University Press},\ \bibinfo {year} {2009})\BibitemShut {NoStop}%
\bibitem [{\citenamefont {Buchdahl}(1959)}]{Buchdahl1959}%
  \BibitemOpen
  \bibfield  {author} {\bibinfo {author} {\bibfnamefont {H.~A.}\ \bibnamefont {Buchdahl}},\ }\bibfield  {title} {\bibinfo {title} {General relativistic fluid spheres},\ }\href {https://doi.org/10.1103/PhysRev.116.1027} {\bibfield  {journal} {\bibinfo  {journal} {Phys. Rev.}\ }\textbf {\bibinfo {volume} {116}},\ \bibinfo {pages} {1027} (\bibinfo {year} {1959})}\BibitemShut {NoStop}%
\bibitem [{\citenamefont {Oppenheimer}\ and\ \citenamefont {Volkoff}(1939)}]{OppenheimerVolkoff1939}%
  \BibitemOpen
  \bibfield  {author} {\bibinfo {author} {\bibfnamefont {J.~R.}\ \bibnamefont {Oppenheimer}}\ and\ \bibinfo {author} {\bibfnamefont {G.~M.}\ \bibnamefont {Volkoff}},\ }\bibfield  {title} {\bibinfo {title} {On massive neutron cores},\ }\href {https://doi.org/10.1103/PhysRev.55.374} {\bibfield  {journal} {\bibinfo  {journal} {Phys. Rev.}\ }\textbf {\bibinfo {volume} {55}},\ \bibinfo {pages} {374} (\bibinfo {year} {1939})}\BibitemShut {NoStop}%
\bibitem [{\citenamefont {Rhoades}\ and\ \citenamefont {Ruffini}(1974)}]{RhoadesRuffini1974}%
  \BibitemOpen
  \bibfield  {author} {\bibinfo {author} {\bibfnamefont {C.~E.}\ \bibnamefont {Rhoades}}\ and\ \bibinfo {author} {\bibfnamefont {R.}~\bibnamefont {Ruffini}},\ }\bibfield  {title} {\bibinfo {title} {Maximum mass of a neutron star},\ }\href {https://doi.org/10.1103/PhysRevLett.32.324} {\bibfield  {journal} {\bibinfo  {journal} {Phys. Rev. Lett.}\ }\textbf {\bibinfo {volume} {32}},\ \bibinfo {pages} {324} (\bibinfo {year} {1974})}\BibitemShut {NoStop}%
\bibitem [{\citenamefont {{Friedman}}\ and\ \citenamefont {{Ipser}}(1987)}]{FriedmanIpser1987}%
  \BibitemOpen
  \bibfield  {author} {\bibinfo {author} {\bibfnamefont {J.~L.}\ \bibnamefont {{Friedman}}}\ and\ \bibinfo {author} {\bibfnamefont {J.~R.}\ \bibnamefont {{Ipser}}},\ }\bibfield  {title} {\bibinfo {title} {{On the Maximum Mass of a Uniformly Rotating Neutron Star}},\ }\href {https://doi.org/10.1086/165088} {\bibfield  {journal} {\bibinfo  {journal} {\apj}\ }\textbf {\bibinfo {volume} {314}},\ \bibinfo {pages} {594} (\bibinfo {year} {1987})}\BibitemShut {NoStop}%
\bibitem [{\citenamefont {Lynden-Bell}\ and\ \citenamefont {Rees}(1971)}]{Lynden-BellRees1971}%
  \BibitemOpen
  \bibfield  {author} {\bibinfo {author} {\bibfnamefont {D.}~\bibnamefont {Lynden-Bell}}\ and\ \bibinfo {author} {\bibfnamefont {M.~J.}\ \bibnamefont {Rees}},\ }\bibfield  {title} {\bibinfo {title} {{On Quasars, Dust and the Galactic Centre}},\ }\href {https://doi.org/10.1093/mnras/152.4.461} {\bibfield  {journal} {\bibinfo  {journal} {Monthly Notices of the Royal Astronomical Society}\ }\textbf {\bibinfo {volume} {152}},\ \bibinfo {pages} {461} (\bibinfo {year} {1971})},\ \Eprint {https://arxiv.org/abs/https://academic.oup.com/mnras/article-pdf/152/4/461/9402718/mnras152-0461.pdf} {https://academic.oup.com/mnras/article-pdf/152/4/461/9402718/mnras152-0461.pdf} \BibitemShut {NoStop}%
\bibitem [{\citenamefont {{Schmidt}}(1963)}]{Schmidt1963}%
  \BibitemOpen
  \bibfield  {author} {\bibinfo {author} {\bibfnamefont {M.}~\bibnamefont {{Schmidt}}},\ }\bibfield  {title} {\bibinfo {title} {{3C 273 : A Star-Like Object with Large Red-Shift}},\ }\href {https://doi.org/10.1038/1971040a0} {\bibfield  {journal} {\bibinfo  {journal} {\nat}\ }\textbf {\bibinfo {volume} {197}},\ \bibinfo {pages} {1040} (\bibinfo {year} {1963})}\BibitemShut {NoStop}%
\bibitem [{\citenamefont {{Bolton}}(1972)}]{Bolton1972}%
  \BibitemOpen
  \bibfield  {author} {\bibinfo {author} {\bibfnamefont {C.~T.}\ \bibnamefont {{Bolton}}},\ }\bibfield  {title} {\bibinfo {title} {{Identification of Cygnus X-1 with HDE 226868}},\ }\href {https://doi.org/10.1038/235271b0} {\bibfield  {journal} {\bibinfo  {journal} {\nat}\ }\textbf {\bibinfo {volume} {235}},\ \bibinfo {pages} {271} (\bibinfo {year} {1972})}\BibitemShut {NoStop}%
\bibitem [{\citenamefont {Ghez}\ \emph {et~al.}(2005)\citenamefont {Ghez}, \citenamefont {Salim}, \citenamefont {Hornstein}, \citenamefont {Tanner}, \citenamefont {Morris}, \citenamefont {Becklin},\ and\ \citenamefont {Duchene}}]{Ghezetal2003}%
  \BibitemOpen
  \bibfield  {author} {\bibinfo {author} {\bibfnamefont {A.~M.}\ \bibnamefont {Ghez}}, \bibinfo {author} {\bibfnamefont {S.}~\bibnamefont {Salim}}, \bibinfo {author} {\bibfnamefont {S.~D.}\ \bibnamefont {Hornstein}}, \bibinfo {author} {\bibfnamefont {A.}~\bibnamefont {Tanner}}, \bibinfo {author} {\bibfnamefont {M.}~\bibnamefont {Morris}}, \bibinfo {author} {\bibfnamefont {E.~E.}\ \bibnamefont {Becklin}},\ and\ \bibinfo {author} {\bibfnamefont {G.}~\bibnamefont {Duchene}},\ }\bibfield  {title} {\bibinfo {title} {{Stellar orbits around the galactic center black hole}},\ }\href {https://doi.org/10.1086/427175} {\bibfield  {journal} {\bibinfo  {journal} {Astrophys. J.}\ }\textbf {\bibinfo {volume} {620}},\ \bibinfo {pages} {744} (\bibinfo {year} {2005})},\ \Eprint {https://arxiv.org/abs/astro-ph/0306130} {arXiv:astro-ph/0306130} \BibitemShut {NoStop}%
\bibitem [{\citenamefont {Abbott}\ \emph {et~al.}(2016)\citenamefont {Abbott} \emph {et~al.}}]{LIGOScientific2016}%
  \BibitemOpen
  \bibfield  {author} {\bibinfo {author} {\bibfnamefont {B.~P.}\ \bibnamefont {Abbott}} \emph {et~al.} (\bibinfo {collaboration} {LIGO Scientific, Virgo}),\ }\bibfield  {title} {\bibinfo {title} {{Observation of Gravitational Waves from a Binary Black Hole Merger}},\ }\href {https://doi.org/10.1103/PhysRevLett.116.061102} {\bibfield  {journal} {\bibinfo  {journal} {Phys. Rev. Lett.}\ }\textbf {\bibinfo {volume} {116}},\ \bibinfo {pages} {061102} (\bibinfo {year} {2016})},\ \Eprint {https://arxiv.org/abs/1602.03837} {arXiv:1602.03837 [gr-qc]} \BibitemShut {NoStop}%
\bibitem [{\citenamefont {Abbott}\ \emph {et~al.}(2017)\citenamefont {Abbott} \emph {et~al.}}]{LIGOScientific2017}%
  \BibitemOpen
  \bibfield  {author} {\bibinfo {author} {\bibfnamefont {B.~P.}\ \bibnamefont {Abbott}} \emph {et~al.} (\bibinfo {collaboration} {LIGO Scientific, Virgo}),\ }\bibfield  {title} {\bibinfo {title} {{GW170817: Observation of Gravitational Waves from a Binary Neutron Star Inspiral}},\ }\href {https://doi.org/10.1103/PhysRevLett.119.161101} {\bibfield  {journal} {\bibinfo  {journal} {Phys. Rev. Lett.}\ }\textbf {\bibinfo {volume} {119}},\ \bibinfo {pages} {161101} (\bibinfo {year} {2017})},\ \Eprint {https://arxiv.org/abs/1710.05832} {arXiv:1710.05832 [gr-qc]} \BibitemShut {NoStop}%
\bibitem [{\citenamefont {Akiyama}\ \emph {et~al.}(2019)\citenamefont {Akiyama} \emph {et~al.}}]{EventHorizonTelescope2019}%
  \BibitemOpen
  \bibfield  {author} {\bibinfo {author} {\bibfnamefont {K.}~\bibnamefont {Akiyama}} \emph {et~al.} (\bibinfo {collaboration} {Event Horizon Telescope}),\ }\bibfield  {title} {\bibinfo {title} {{First M87 Event Horizon Telescope Results. I. The Shadow of the Supermassive Black Hole}},\ }\href {https://doi.org/10.3847/2041-8213/ab0ec7} {\bibfield  {journal} {\bibinfo  {journal} {Astrophys. J. Lett.}\ }\textbf {\bibinfo {volume} {875}},\ \bibinfo {pages} {L1} (\bibinfo {year} {2019})},\ \Eprint {https://arxiv.org/abs/1906.11238} {arXiv:1906.11238 [astro-ph.GA]} \BibitemShut {NoStop}%
\bibitem [{\citenamefont {Akiyama}\ \emph {et~al.}(2022)\citenamefont {Akiyama} \emph {et~al.}}]{EventHorizonTelescope2022}%
  \BibitemOpen
  \bibfield  {author} {\bibinfo {author} {\bibfnamefont {K.}~\bibnamefont {Akiyama}} \emph {et~al.} (\bibinfo {collaboration} {Event Horizon Telescope}),\ }\bibfield  {title} {\bibinfo {title} {{First Sagittarius A* Event Horizon Telescope Results. I. The Shadow of the Supermassive Black Hole in the Center of the Milky Way}},\ }\href {https://doi.org/10.3847/2041-8213/ac6674} {\bibfield  {journal} {\bibinfo  {journal} {Astrophys. J. Lett.}\ }\textbf {\bibinfo {volume} {930}},\ \bibinfo {pages} {L12} (\bibinfo {year} {2022})}\BibitemShut {NoStop}%
\bibitem [{\citenamefont {Carballo-Rubio}\ \emph {et~al.}(2018)\citenamefont {Carballo-Rubio}, \citenamefont {Di~Filippo}, \citenamefont {Liberati},\ and\ \citenamefont {Visser}}]{CarballoRubioetal2018}%
  \BibitemOpen
  \bibfield  {author} {\bibinfo {author} {\bibfnamefont {R.}~\bibnamefont {Carballo-Rubio}}, \bibinfo {author} {\bibfnamefont {F.}~\bibnamefont {Di~Filippo}}, \bibinfo {author} {\bibfnamefont {S.}~\bibnamefont {Liberati}},\ and\ \bibinfo {author} {\bibfnamefont {M.}~\bibnamefont {Visser}},\ }\bibfield  {title} {\bibinfo {title} {{Phenomenological aspects of black holes beyond general relativity}},\ }\href {https://doi.org/10.1103/PhysRevD.98.124009} {\bibfield  {journal} {\bibinfo  {journal} {Phys. Rev. D}\ }\textbf {\bibinfo {volume} {98}},\ \bibinfo {pages} {124009} (\bibinfo {year} {2018})},\ \Eprint {https://arxiv.org/abs/1809.08238} {arXiv:1809.08238 [gr-qc]} \BibitemShut {NoStop}%
\bibitem [{\citenamefont {{Bardeen}}(1968)}]{1968qtr..conf...87B}%
  \BibitemOpen
  \bibfield  {author} {\bibinfo {author} {\bibfnamefont {J.}~\bibnamefont {{Bardeen}}},\ }\bibfield  {title} {\bibinfo {title} {{Non-singular general relativistic gravitational collapse}},\ }in\ \href@noop {} {\emph {\bibinfo {booktitle} {Proceedings of the 5th International Conference on Gravitation and the Theory of Relativity}}}\ (\bibinfo {year} {1968})\ p.~\bibinfo {pages} {87}\BibitemShut {NoStop}%
\bibitem [{\citenamefont {Hayward}(1993)}]{Hayward1993}%
  \BibitemOpen
  \bibfield  {author} {\bibinfo {author} {\bibfnamefont {S.~A.}\ \bibnamefont {Hayward}},\ }\bibfield  {title} {\bibinfo {title} {{Marginal surfaces and apparent horizons}},\ }\href@noop {} {\  (\bibinfo {year} {1993})},\ \Eprint {https://arxiv.org/abs/gr-qc/9303006} {arXiv:gr-qc/9303006} \BibitemShut {NoStop}%
\bibitem [{\citenamefont {Simpson}\ and\ \citenamefont {Visser}(2019)}]{SimpsonVisser2018}%
  \BibitemOpen
  \bibfield  {author} {\bibinfo {author} {\bibfnamefont {A.}~\bibnamefont {Simpson}}\ and\ \bibinfo {author} {\bibfnamefont {M.}~\bibnamefont {Visser}},\ }\bibfield  {title} {\bibinfo {title} {{Black-bounce to traversable wormhole}},\ }\href {https://doi.org/10.1088/1475-7516/2019/02/042} {\bibfield  {journal} {\bibinfo  {journal} {JCAP}\ }\textbf {\bibinfo {volume} {02}},\ \bibinfo {pages} {042}},\ \Eprint {https://arxiv.org/abs/1812.07114} {arXiv:1812.07114 [gr-qc]} \BibitemShut {NoStop}%
\bibitem [{\citenamefont {Franzin}\ \emph {et~al.}(2024)\citenamefont {Franzin}, \citenamefont {Liberati},\ and\ \citenamefont {Vellucci}}]{Franzin:2023slm}%
  \BibitemOpen
  \bibfield  {author} {\bibinfo {author} {\bibfnamefont {E.}~\bibnamefont {Franzin}}, \bibinfo {author} {\bibfnamefont {S.}~\bibnamefont {Liberati}},\ and\ \bibinfo {author} {\bibfnamefont {V.}~\bibnamefont {Vellucci}},\ }\bibfield  {title} {\bibinfo {title} {{From regular black holes to horizonless objects: quasi-normal modes, instabilities and spectroscopy}},\ }\href {https://doi.org/10.1088/1475-7516/2024/01/020} {\bibfield  {journal} {\bibinfo  {journal} {JCAP}\ }\textbf {\bibinfo {volume} {01}},\ \bibinfo {pages} {020}},\ \Eprint {https://arxiv.org/abs/2310.11990} {arXiv:2310.11990 [gr-qc]} \BibitemShut {NoStop}%
\bibitem [{\citenamefont {Mazur}\ and\ \citenamefont {Mottola}(2004)}]{MazurMottola2004}%
  \BibitemOpen
  \bibfield  {author} {\bibinfo {author} {\bibfnamefont {P.~O.}\ \bibnamefont {Mazur}}\ and\ \bibinfo {author} {\bibfnamefont {E.}~\bibnamefont {Mottola}},\ }\bibfield  {title} {\bibinfo {title} {{Gravitational vacuum condensate stars}},\ }\href {https://doi.org/10.1073/pnas.0402717101} {\bibfield  {journal} {\bibinfo  {journal} {Proc. Nat. Acad. Sci.}\ }\textbf {\bibinfo {volume} {101}},\ \bibinfo {pages} {9545} (\bibinfo {year} {2004})},\ \Eprint {https://arxiv.org/abs/gr-qc/0407075} {arXiv:gr-qc/0407075} \BibitemShut {NoStop}%
\bibitem [{\citenamefont {Mathur}(2005)}]{Mathur2005}%
  \BibitemOpen
  \bibfield  {author} {\bibinfo {author} {\bibfnamefont {S.~D.}\ \bibnamefont {Mathur}},\ }\bibfield  {title} {\bibinfo {title} {{The Fuzzball proposal for black holes: An Elementary review}},\ }\bibfield  {booktitle} {\emph {\bibinfo {booktitle} {{The quantum structure of space-time and the geometric nature of fundamental interactions. Proceedings, 4th Meeting, RTN2004, Kolymbari, Crete, Greece, September 5-10, 2004}}},\ }\href {https://doi.org/10.1002/prop.200410203} {\bibfield  {journal} {\bibinfo  {journal} {Fortsch. Phys.}\ }\textbf {\bibinfo {volume} {53}},\ \bibinfo {pages} {793} (\bibinfo {year} {2005})},\ \Eprint {https://arxiv.org/abs/hep-th/0502050} {arXiv:hep-th/0502050 [hep-th]} \BibitemShut {NoStop}%
%%CITATION = HEP-TH/0502050;%%
\bibitem [{\citenamefont {Ikeda}\ \emph {et~al.}(2021)\citenamefont {Ikeda}, \citenamefont {Bianchi}, \citenamefont {Consoli}, \citenamefont {Grillo}, \citenamefont {Morales}, \citenamefont {Pani},\ and\ \citenamefont {Raposo}}]{Ikedaetal2021}%
  \BibitemOpen
  \bibfield  {author} {\bibinfo {author} {\bibfnamefont {T.}~\bibnamefont {Ikeda}}, \bibinfo {author} {\bibfnamefont {M.}~\bibnamefont {Bianchi}}, \bibinfo {author} {\bibfnamefont {D.}~\bibnamefont {Consoli}}, \bibinfo {author} {\bibfnamefont {A.}~\bibnamefont {Grillo}}, \bibinfo {author} {\bibfnamefont {J.~F.}\ \bibnamefont {Morales}}, \bibinfo {author} {\bibfnamefont {P.}~\bibnamefont {Pani}},\ and\ \bibinfo {author} {\bibfnamefont {G.}~\bibnamefont {Raposo}},\ }\bibfield  {title} {\bibinfo {title} {Black-hole microstate spectroscopy: Ringdown, quasinormal modes, and echoes},\ }\href {https://doi.org/10.1103/PhysRevD.104.066021} {\bibfield  {journal} {\bibinfo  {journal} {Phys. Rev. D}\ }\textbf {\bibinfo {volume} {104}},\ \bibinfo {pages} {066021} (\bibinfo {year} {2021})}\BibitemShut {NoStop}%
\bibitem [{\citenamefont {Brustein}\ and\ \citenamefont {Medved}(2017)}]{BrusteinMedved2017}%
  \BibitemOpen
  \bibfield  {author} {\bibinfo {author} {\bibfnamefont {R.}~\bibnamefont {Brustein}}\ and\ \bibinfo {author} {\bibfnamefont {A.~J.~M.}\ \bibnamefont {Medved}},\ }\bibfield  {title} {\bibinfo {title} {{Black holes as collapsed polymers}},\ }\href {https://doi.org/10.1002/prop.201600114} {\bibfield  {journal} {\bibinfo  {journal} {Fortsch. Phys.}\ }\textbf {\bibinfo {volume} {65}},\ \bibinfo {pages} {1600114} (\bibinfo {year} {2017})},\ \Eprint {https://arxiv.org/abs/1602.07706} {arXiv:1602.07706 [hep-th]} \BibitemShut {NoStop}%
\bibitem [{\citenamefont {Cardoso}\ and\ \citenamefont {Pani}(2017{\natexlab{a}})}]{CardosoPani2017}%
  \BibitemOpen
  \bibfield  {author} {\bibinfo {author} {\bibfnamefont {V.}~\bibnamefont {Cardoso}}\ and\ \bibinfo {author} {\bibfnamefont {P.}~\bibnamefont {Pani}},\ }\bibfield  {title} {\bibinfo {title} {{Tests for the existence of black holes through gravitational wave echoes}},\ }\href {https://doi.org/https://doi.org/10.1038/s41550-017-0225-y} {\bibfield  {journal} {\bibinfo  {journal} {Nat. Astron.}\ }\textbf {\bibinfo {volume} {1}},\ \bibinfo {pages} {586} (\bibinfo {year} {2017}{\natexlab{a}})},\ \Eprint {https://arxiv.org/abs/1707.03021} {arXiv:1707.03021 [gr-qc]} \BibitemShut {NoStop}%
\bibitem [{\citenamefont {Raposo}\ \emph {et~al.}(2019{\natexlab{a}})\citenamefont {Raposo}, \citenamefont {Pani}, \citenamefont {Bezares}, \citenamefont {Palenzuela},\ and\ \citenamefont {Cardoso}}]{Raposoetal2018}%
  \BibitemOpen
  \bibfield  {author} {\bibinfo {author} {\bibfnamefont {G.}~\bibnamefont {Raposo}}, \bibinfo {author} {\bibfnamefont {P.}~\bibnamefont {Pani}}, \bibinfo {author} {\bibfnamefont {M.}~\bibnamefont {Bezares}}, \bibinfo {author} {\bibfnamefont {C.}~\bibnamefont {Palenzuela}},\ and\ \bibinfo {author} {\bibfnamefont {V.}~\bibnamefont {Cardoso}},\ }\bibfield  {title} {\bibinfo {title} {{Anisotropic stars as ultracompact objects in General Relativity}},\ }\href {https://doi.org/10.1103/PhysRevD.99.104072} {\bibfield  {journal} {\bibinfo  {journal} {Phys. Rev.}\ }\textbf {\bibinfo {volume} {D99}},\ \bibinfo {pages} {104072} (\bibinfo {year} {2019}{\natexlab{a}})},\ \Eprint {https://arxiv.org/abs/1811.07917} {arXiv:1811.07917 [gr-qc]} \BibitemShut {NoStop}%
%%CITATION = ARXIV:1811.07917;%%
\bibitem [{\citenamefont {Holdom}\ and\ \citenamefont {Ren}(2017)}]{HoldomRen2017}%
  \BibitemOpen
  \bibfield  {author} {\bibinfo {author} {\bibfnamefont {B.}~\bibnamefont {Holdom}}\ and\ \bibinfo {author} {\bibfnamefont {J.}~\bibnamefont {Ren}},\ }\bibfield  {title} {\bibinfo {title} {Not quite a black hole},\ }\href {https://doi.org/10.1103/PhysRevD.95.084034} {\bibfield  {journal} {\bibinfo  {journal} {Phys. Rev. D}\ }\textbf {\bibinfo {volume} {95}},\ \bibinfo {pages} {084034} (\bibinfo {year} {2017})}\BibitemShut {NoStop}%
\bibitem [{\citenamefont {Pani}\ \emph {et~al.}(2010)\citenamefont {Pani}, \citenamefont {Berti}, \citenamefont {Cardoso}, \citenamefont {Chen},\ and\ \citenamefont {Norte}}]{Pani:2010em}%
  \BibitemOpen
  \bibfield  {author} {\bibinfo {author} {\bibfnamefont {P.}~\bibnamefont {Pani}}, \bibinfo {author} {\bibfnamefont {E.}~\bibnamefont {Berti}}, \bibinfo {author} {\bibfnamefont {V.}~\bibnamefont {Cardoso}}, \bibinfo {author} {\bibfnamefont {Y.}~\bibnamefont {Chen}},\ and\ \bibinfo {author} {\bibfnamefont {R.}~\bibnamefont {Norte}},\ }\bibfield  {title} {\bibinfo {title} {{Gravitational-wave signatures of the absence of an event horizon. II. Extreme mass ratio inspirals in the spacetime of a thin-shell gravastar}},\ }\href {https://doi.org/10.1103/PhysRevD.81.084011} {\bibfield  {journal} {\bibinfo  {journal} {Phys. Rev. D}\ }\textbf {\bibinfo {volume} {81}},\ \bibinfo {pages} {084011} (\bibinfo {year} {2010})},\ \Eprint {https://arxiv.org/abs/1001.3031} {arXiv:1001.3031 [gr-qc]} \BibitemShut {NoStop}%
\bibitem [{\citenamefont {Kesden}\ \emph {et~al.}(2005)\citenamefont {Kesden}, \citenamefont {Gair},\ and\ \citenamefont {Kamionkowski}}]{Kesden:2004qx}%
  \BibitemOpen
  \bibfield  {author} {\bibinfo {author} {\bibfnamefont {M.}~\bibnamefont {Kesden}}, \bibinfo {author} {\bibfnamefont {J.}~\bibnamefont {Gair}},\ and\ \bibinfo {author} {\bibfnamefont {M.}~\bibnamefont {Kamionkowski}},\ }\bibfield  {title} {\bibinfo {title} {{Gravitational-wave signature of an inspiral into a supermassive horizonless object}},\ }\href {https://doi.org/10.1103/PhysRevD.71.044015} {\bibfield  {journal} {\bibinfo  {journal} {Phys. Rev. D}\ }\textbf {\bibinfo {volume} {71}},\ \bibinfo {pages} {044015} (\bibinfo {year} {2005})},\ \Eprint {https://arxiv.org/abs/astro-ph/0411478} {arXiv:astro-ph/0411478} \BibitemShut {NoStop}%
\bibitem [{\citenamefont {Cardoso}\ \emph {et~al.}(2017)\citenamefont {Cardoso}, \citenamefont {Franzin}, \citenamefont {Maselli}, \citenamefont {Pani},\ and\ \citenamefont {Raposo}}]{Cardoso:2017cfl}%
  \BibitemOpen
  \bibfield  {author} {\bibinfo {author} {\bibfnamefont {V.}~\bibnamefont {Cardoso}}, \bibinfo {author} {\bibfnamefont {E.}~\bibnamefont {Franzin}}, \bibinfo {author} {\bibfnamefont {A.}~\bibnamefont {Maselli}}, \bibinfo {author} {\bibfnamefont {P.}~\bibnamefont {Pani}},\ and\ \bibinfo {author} {\bibfnamefont {G.}~\bibnamefont {Raposo}},\ }\bibfield  {title} {\bibinfo {title} {{Testing strong-field gravity with tidal Love numbers}},\ }\href {https://doi.org/10.1103/PhysRevD.95.084014} {\bibfield  {journal} {\bibinfo  {journal} {Phys. Rev. D}\ }\textbf {\bibinfo {volume} {95}},\ \bibinfo {pages} {084014} (\bibinfo {year} {2017})},\ \bibinfo {note} {[Addendum: Phys.Rev.D 95, 089901 (2017)]},\ \Eprint {https://arxiv.org/abs/1701.01116} {arXiv:1701.01116 [gr-qc]} \BibitemShut {NoStop}%
\bibitem [{\citenamefont {Herdeiro}\ \emph {et~al.}(2020)\citenamefont {Herdeiro}, \citenamefont {Panotopoulos},\ and\ \citenamefont {Radu}}]{Herdeiro:2020kba}%
  \BibitemOpen
  \bibfield  {author} {\bibinfo {author} {\bibfnamefont {C.~A.~R.}\ \bibnamefont {Herdeiro}}, \bibinfo {author} {\bibfnamefont {G.}~\bibnamefont {Panotopoulos}},\ and\ \bibinfo {author} {\bibfnamefont {E.}~\bibnamefont {Radu}},\ }\bibfield  {title} {\bibinfo {title} {{Tidal Love numbers of Proca stars}},\ }\href {https://doi.org/10.1088/1475-7516/2020/08/029} {\bibfield  {journal} {\bibinfo  {journal} {JCAP}\ }\textbf {\bibinfo {volume} {08}},\ \bibinfo {pages} {029}},\ \Eprint {https://arxiv.org/abs/2006.11083} {arXiv:2006.11083 [gr-qc]} \BibitemShut {NoStop}%
\bibitem [{\citenamefont {Di~Filippo}(2024)}]{DiFilippo:2024mnc}%
  \BibitemOpen
  \bibfield  {author} {\bibinfo {author} {\bibfnamefont {F.}~\bibnamefont {Di~Filippo}},\ }\bibfield  {title} {\bibinfo {title} {{On the nature of inner light-rings}},\ }\href@noop {} {\  (\bibinfo {year} {2024})},\ \Eprint {https://arxiv.org/abs/2404.07357} {arXiv:2404.07357 [gr-qc]} \BibitemShut {NoStop}%
\bibitem [{\citenamefont {Cardoso}\ \emph {et~al.}(2016)\citenamefont {Cardoso}, \citenamefont {Franzin},\ and\ \citenamefont {Pani}}]{Cardoso:2016rao}%
  \BibitemOpen
  \bibfield  {author} {\bibinfo {author} {\bibfnamefont {V.}~\bibnamefont {Cardoso}}, \bibinfo {author} {\bibfnamefont {E.}~\bibnamefont {Franzin}},\ and\ \bibinfo {author} {\bibfnamefont {P.}~\bibnamefont {Pani}},\ }\bibfield  {title} {\bibinfo {title} {{Is the gravitational-wave ringdown a probe of the event horizon?}},\ }\href {https://doi.org/10.1103/PhysRevLett.116.171101} {\bibfield  {journal} {\bibinfo  {journal} {Phys. Rev. Lett.}\ }\textbf {\bibinfo {volume} {116}},\ \bibinfo {pages} {171101} (\bibinfo {year} {2016})},\ \bibinfo {note} {[Erratum: Phys.Rev.Lett. 117, 089902 (2016)]},\ \Eprint {https://arxiv.org/abs/1602.07309} {arXiv:1602.07309 [gr-qc]} \BibitemShut {NoStop}%
\bibitem [{\citenamefont {Vellucci}\ \emph {et~al.}(2023)\citenamefont {Vellucci}, \citenamefont {Franzin},\ and\ \citenamefont {Liberati}}]{Vellucci:2022hpl}%
  \BibitemOpen
  \bibfield  {author} {\bibinfo {author} {\bibfnamefont {V.}~\bibnamefont {Vellucci}}, \bibinfo {author} {\bibfnamefont {E.}~\bibnamefont {Franzin}},\ and\ \bibinfo {author} {\bibfnamefont {S.}~\bibnamefont {Liberati}},\ }\bibfield  {title} {\bibinfo {title} {{Echoes from backreacting exotic compact objects}},\ }\href {https://doi.org/10.1103/PhysRevD.107.044027} {\bibfield  {journal} {\bibinfo  {journal} {Phys. Rev. D}\ }\textbf {\bibinfo {volume} {107}},\ \bibinfo {pages} {044027} (\bibinfo {year} {2023})},\ \Eprint {https://arxiv.org/abs/2205.14170} {arXiv:2205.14170 [gr-qc]} \BibitemShut {NoStop}%
\bibitem [{\citenamefont {Abedi}\ \emph {et~al.}(2017{\natexlab{a}})\citenamefont {Abedi}, \citenamefont {Dykaar},\ and\ \citenamefont {Afshordi}}]{Abedi:2016hgu}%
  \BibitemOpen
  \bibfield  {author} {\bibinfo {author} {\bibfnamefont {J.}~\bibnamefont {Abedi}}, \bibinfo {author} {\bibfnamefont {H.}~\bibnamefont {Dykaar}},\ and\ \bibinfo {author} {\bibfnamefont {N.}~\bibnamefont {Afshordi}},\ }\bibfield  {title} {\bibinfo {title} {{Echoes from the Abyss: Tentative evidence for Planck-scale structure at black hole horizons}},\ }\href {https://doi.org/10.1103/PhysRevD.96.082004} {\bibfield  {journal} {\bibinfo  {journal} {Phys. Rev. D}\ }\textbf {\bibinfo {volume} {96}},\ \bibinfo {pages} {082004} (\bibinfo {year} {2017}{\natexlab{a}})},\ \Eprint {https://arxiv.org/abs/1612.00266} {arXiv:1612.00266 [gr-qc]} \BibitemShut {NoStop}%
\bibitem [{\citenamefont {Conklin}\ \emph {et~al.}(2018)\citenamefont {Conklin}, \citenamefont {Holdom},\ and\ \citenamefont {Ren}}]{Conklin:2017lwb}%
  \BibitemOpen
  \bibfield  {author} {\bibinfo {author} {\bibfnamefont {R.~S.}\ \bibnamefont {Conklin}}, \bibinfo {author} {\bibfnamefont {B.}~\bibnamefont {Holdom}},\ and\ \bibinfo {author} {\bibfnamefont {J.}~\bibnamefont {Ren}},\ }\bibfield  {title} {\bibinfo {title} {{Gravitational wave echoes through new windows}},\ }\href {https://doi.org/10.1103/PhysRevD.98.044021} {\bibfield  {journal} {\bibinfo  {journal} {Phys. Rev. D}\ }\textbf {\bibinfo {volume} {98}},\ \bibinfo {pages} {044021} (\bibinfo {year} {2018})},\ \Eprint {https://arxiv.org/abs/1712.06517} {arXiv:1712.06517 [gr-qc]} \BibitemShut {NoStop}%
\bibitem [{\citenamefont {Abedi}\ and\ \citenamefont {Afshordi}(2019)}]{Abedi:2018npz}%
  \BibitemOpen
  \bibfield  {author} {\bibinfo {author} {\bibfnamefont {J.}~\bibnamefont {Abedi}}\ and\ \bibinfo {author} {\bibfnamefont {N.}~\bibnamefont {Afshordi}},\ }\bibfield  {title} {\bibinfo {title} {{Echoes from the Abyss: A highly spinning black hole remnant for the binary neutron star merger GW170817}},\ }\href {https://doi.org/10.1088/1475-7516/2019/11/010} {\bibfield  {journal} {\bibinfo  {journal} {JCAP}\ }\textbf {\bibinfo {volume} {11}},\ \bibinfo {pages} {010}},\ \Eprint {https://arxiv.org/abs/1803.10454} {arXiv:1803.10454 [gr-qc]} \BibitemShut {NoStop}%
\bibitem [{\citenamefont {Westerweck}\ \emph {et~al.}(2018)\citenamefont {Westerweck}, \citenamefont {Nielsen}, \citenamefont {Fischer-Birnholtz}, \citenamefont {Cabero}, \citenamefont {Capano}, \citenamefont {Dent}, \citenamefont {Krishnan}, \citenamefont {Meadors},\ and\ \citenamefont {Nitz}}]{Westerweck:2017hus}%
  \BibitemOpen
  \bibfield  {author} {\bibinfo {author} {\bibfnamefont {J.}~\bibnamefont {Westerweck}}, \bibinfo {author} {\bibfnamefont {A.}~\bibnamefont {Nielsen}}, \bibinfo {author} {\bibfnamefont {O.}~\bibnamefont {Fischer-Birnholtz}}, \bibinfo {author} {\bibfnamefont {M.}~\bibnamefont {Cabero}}, \bibinfo {author} {\bibfnamefont {C.}~\bibnamefont {Capano}}, \bibinfo {author} {\bibfnamefont {T.}~\bibnamefont {Dent}}, \bibinfo {author} {\bibfnamefont {B.}~\bibnamefont {Krishnan}}, \bibinfo {author} {\bibfnamefont {G.}~\bibnamefont {Meadors}},\ and\ \bibinfo {author} {\bibfnamefont {A.~H.}\ \bibnamefont {Nitz}},\ }\bibfield  {title} {\bibinfo {title} {{Low significance of evidence for black hole echoes in gravitational wave data}},\ }\href {https://doi.org/10.1103/PhysRevD.97.124037} {\bibfield  {journal} {\bibinfo  {journal} {Phys. Rev. D}\ }\textbf {\bibinfo {volume} {97}},\ \bibinfo {pages} {124037} (\bibinfo {year} {2018})},\ \Eprint {https://arxiv.org/abs/1712.09966} {arXiv:1712.09966 [gr-qc]} \BibitemShut {NoStop}%
\bibitem [{\citenamefont {Nielsen}\ \emph {et~al.}(2019)\citenamefont {Nielsen}, \citenamefont {Capano}, \citenamefont {Birnholtz},\ and\ \citenamefont {Westerweck}}]{Nielsen:2018lkf}%
  \BibitemOpen
  \bibfield  {author} {\bibinfo {author} {\bibfnamefont {A.~B.}\ \bibnamefont {Nielsen}}, \bibinfo {author} {\bibfnamefont {C.~D.}\ \bibnamefont {Capano}}, \bibinfo {author} {\bibfnamefont {O.}~\bibnamefont {Birnholtz}},\ and\ \bibinfo {author} {\bibfnamefont {J.}~\bibnamefont {Westerweck}},\ }\bibfield  {title} {\bibinfo {title} {{Parameter estimation and statistical significance of echoes following black hole signals in the first Advanced LIGO observing run}},\ }\href {https://doi.org/10.1103/PhysRevD.99.104012} {\bibfield  {journal} {\bibinfo  {journal} {Phys. Rev. D}\ }\textbf {\bibinfo {volume} {99}},\ \bibinfo {pages} {104012} (\bibinfo {year} {2019})},\ \Eprint {https://arxiv.org/abs/1811.04904} {arXiv:1811.04904 [gr-qc]} \BibitemShut {NoStop}%
\bibitem [{\citenamefont {Lo}\ \emph {et~al.}(2019)\citenamefont {Lo}, \citenamefont {Li},\ and\ \citenamefont {Weinstein}}]{Lo:2018sep}%
  \BibitemOpen
  \bibfield  {author} {\bibinfo {author} {\bibfnamefont {R.~K.~L.}\ \bibnamefont {Lo}}, \bibinfo {author} {\bibfnamefont {T.~G.~F.}\ \bibnamefont {Li}},\ and\ \bibinfo {author} {\bibfnamefont {A.~J.}\ \bibnamefont {Weinstein}},\ }\bibfield  {title} {\bibinfo {title} {{Template-based Gravitational-Wave Echoes Search Using Bayesian Model Selection}},\ }\href {https://doi.org/10.1103/PhysRevD.99.084052} {\bibfield  {journal} {\bibinfo  {journal} {Phys. Rev. D}\ }\textbf {\bibinfo {volume} {99}},\ \bibinfo {pages} {084052} (\bibinfo {year} {2019})},\ \Eprint {https://arxiv.org/abs/1811.07431} {arXiv:1811.07431 [gr-qc]} \BibitemShut {NoStop}%
\bibitem [{\citenamefont {Uchikata}\ \emph {et~al.}(2019)\citenamefont {Uchikata}, \citenamefont {Nakano}, \citenamefont {Narikawa}, \citenamefont {Sago}, \citenamefont {Tagoshi},\ and\ \citenamefont {Tanaka}}]{Uchikata:2019frs}%
  \BibitemOpen
  \bibfield  {author} {\bibinfo {author} {\bibfnamefont {N.}~\bibnamefont {Uchikata}}, \bibinfo {author} {\bibfnamefont {H.}~\bibnamefont {Nakano}}, \bibinfo {author} {\bibfnamefont {T.}~\bibnamefont {Narikawa}}, \bibinfo {author} {\bibfnamefont {N.}~\bibnamefont {Sago}}, \bibinfo {author} {\bibfnamefont {H.}~\bibnamefont {Tagoshi}},\ and\ \bibinfo {author} {\bibfnamefont {T.}~\bibnamefont {Tanaka}},\ }\bibfield  {title} {\bibinfo {title} {{Searching for black hole echoes from the LIGO-Virgo Catalog GWTC-1}},\ }\href {https://doi.org/10.1103/PhysRevD.100.062006} {\bibfield  {journal} {\bibinfo  {journal} {Phys. Rev. D}\ }\textbf {\bibinfo {volume} {100}},\ \bibinfo {pages} {062006} (\bibinfo {year} {2019})},\ \Eprint {https://arxiv.org/abs/1906.00838} {arXiv:1906.00838 [gr-qc]} \BibitemShut {NoStop}%
\bibitem [{\citenamefont {Tsang}\ \emph {et~al.}(2020)\citenamefont {Tsang}, \citenamefont {Ghosh}, \citenamefont {Samajdar}, \citenamefont {Chatziioannou}, \citenamefont {Mastrogiovanni}, \citenamefont {Agathos},\ and\ \citenamefont {Van Den~Broeck}}]{Tsang:2019zra}%
  \BibitemOpen
  \bibfield  {author} {\bibinfo {author} {\bibfnamefont {K.~W.}\ \bibnamefont {Tsang}}, \bibinfo {author} {\bibfnamefont {A.}~\bibnamefont {Ghosh}}, \bibinfo {author} {\bibfnamefont {A.}~\bibnamefont {Samajdar}}, \bibinfo {author} {\bibfnamefont {K.}~\bibnamefont {Chatziioannou}}, \bibinfo {author} {\bibfnamefont {S.}~\bibnamefont {Mastrogiovanni}}, \bibinfo {author} {\bibfnamefont {M.}~\bibnamefont {Agathos}},\ and\ \bibinfo {author} {\bibfnamefont {C.}~\bibnamefont {Van Den~Broeck}},\ }\bibfield  {title} {\bibinfo {title} {{A morphology-independent search for gravitational wave echoes in data from the first and second observing runs of Advanced LIGO and Advanced Virgo}},\ }\href {https://doi.org/10.1103/PhysRevD.101.064012} {\bibfield  {journal} {\bibinfo  {journal} {Phys. Rev. D}\ }\textbf {\bibinfo {volume} {101}},\ \bibinfo {pages} {064012} (\bibinfo {year} {2020})},\ \Eprint {https://arxiv.org/abs/1906.11168} {arXiv:1906.11168 [gr-qc]} \BibitemShut {NoStop}%
\bibitem [{\citenamefont {Abbott}\ \emph {et~al.}(2021{\natexlab{a}})\citenamefont {Abbott} \emph {et~al.}}]{LIGOScientific:2020tif}%
  \BibitemOpen
  \bibfield  {author} {\bibinfo {author} {\bibfnamefont {R.}~\bibnamefont {Abbott}} \emph {et~al.} (\bibinfo {collaboration} {LIGO Scientific, Virgo}),\ }\bibfield  {title} {\bibinfo {title} {{Tests of general relativity with binary black holes from the second LIGO-Virgo gravitational-wave transient catalog}},\ }\href {https://doi.org/10.1103/PhysRevD.103.122002} {\bibfield  {journal} {\bibinfo  {journal} {Phys. Rev. D}\ }\textbf {\bibinfo {volume} {103}},\ \bibinfo {pages} {122002} (\bibinfo {year} {2021}{\natexlab{a}})},\ \Eprint {https://arxiv.org/abs/2010.14529} {arXiv:2010.14529 [gr-qc]} \BibitemShut {NoStop}%
\bibitem [{\citenamefont {Abbott}\ \emph {et~al.}(2021{\natexlab{b}})\citenamefont {Abbott} \emph {et~al.}}]{LIGOScientific:2021sio}%
  \BibitemOpen
  \bibfield  {author} {\bibinfo {author} {\bibfnamefont {R.}~\bibnamefont {Abbott}} \emph {et~al.} (\bibinfo {collaboration} {LIGO Scientific, VIRGO, KAGRA}),\ }\bibfield  {title} {\bibinfo {title} {{Tests of General Relativity with GWTC-3}},\ }\href@noop {} {\  (\bibinfo {year} {2021}{\natexlab{b}})},\ \Eprint {https://arxiv.org/abs/2112.06861} {arXiv:2112.06861 [gr-qc]} \BibitemShut {NoStop}%
\bibitem [{\citenamefont {Punturo}\ \emph {et~al.}(2010)\citenamefont {Punturo} \emph {et~al.}}]{Punturo:2010zz}%
  \BibitemOpen
  \bibfield  {author} {\bibinfo {author} {\bibfnamefont {M.}~\bibnamefont {Punturo}} \emph {et~al.},\ }\bibfield  {title} {\bibinfo {title} {{The Einstein Telescope: A third-generation gravitational wave observatory}},\ }\href {https://doi.org/10.1088/0264-9381/27/19/194002} {\bibfield  {journal} {\bibinfo  {journal} {Class. Quantum Grav.}\ }\textbf {\bibinfo {volume} {27}},\ \bibinfo {pages} {194002} (\bibinfo {year} {2010})}\BibitemShut {NoStop}%
\bibitem [{\citenamefont {Branchesi}\ \emph {et~al.}(2023)\citenamefont {Branchesi} \emph {et~al.}}]{Branchesi:2023mws}%
  \BibitemOpen
  \bibfield  {author} {\bibinfo {author} {\bibfnamefont {M.}~\bibnamefont {Branchesi}} \emph {et~al.},\ }\bibfield  {title} {\bibinfo {title} {{Science with the Einstein Telescope: a comparison of different designs}},\ }\href {https://doi.org/10.1088/1475-7516/2023/07/068} {\bibfield  {journal} {\bibinfo  {journal} {JCAP}\ }\textbf {\bibinfo {volume} {07}},\ \bibinfo {pages} {068}},\ \Eprint {https://arxiv.org/abs/2303.15923} {arXiv:2303.15923 [gr-qc]} \BibitemShut {NoStop}%
\bibitem [{\citenamefont {Kalogera}\ \emph {et~al.}(2021)\citenamefont {Kalogera} \emph {et~al.}}]{Kalogera:2021bya}%
  \BibitemOpen
  \bibfield  {author} {\bibinfo {author} {\bibfnamefont {V.}~\bibnamefont {Kalogera}} \emph {et~al.},\ }\bibfield  {title} {\bibinfo {title} {{The Next Generation Global Gravitational Wave Observatory: The Science Book}},\ }\href@noop {} {\  (\bibinfo {year} {2021})},\ \Eprint {https://arxiv.org/abs/2111.06990} {arXiv:2111.06990 [gr-qc]} \BibitemShut {NoStop}%
\bibitem [{\citenamefont {Maggiore}\ \emph {et~al.}(2020)\citenamefont {Maggiore} \emph {et~al.}}]{Maggiore:2019uih}%
  \BibitemOpen
  \bibfield  {author} {\bibinfo {author} {\bibfnamefont {M.}~\bibnamefont {Maggiore}} \emph {et~al.},\ }\bibfield  {title} {\bibinfo {title} {{Science Case for the Einstein Telescope}},\ }\href {https://doi.org/10.1088/1475-7516/2020/03/050} {\bibfield  {journal} {\bibinfo  {journal} {JCAP}\ }\textbf {\bibinfo {volume} {03}},\ \bibinfo {pages} {050}},\ \Eprint {https://arxiv.org/abs/1912.02622} {arXiv:1912.02622 [astro-ph.CO]} \BibitemShut {NoStop}%
\bibitem [{\citenamefont {Reitze}\ \emph {et~al.}(2019)\citenamefont {Reitze} \emph {et~al.}}]{Reitze:2019iox}%
  \BibitemOpen
  \bibfield  {author} {\bibinfo {author} {\bibfnamefont {D.}~\bibnamefont {Reitze}} \emph {et~al.},\ }\bibfield  {title} {\bibinfo {title} {{Cosmic Explorer: The U.S. Contribution to Gravitational-Wave Astronomy beyond LIGO}},\ }\href@noop {} {\bibfield  {journal} {\bibinfo  {journal} {Bull. Am. Astron. Soc.}\ }\textbf {\bibinfo {volume} {51}},\ \bibinfo {pages} {035} (\bibinfo {year} {2019})},\ \Eprint {https://arxiv.org/abs/1907.04833} {arXiv:1907.04833 [astro-ph.IM]} \BibitemShut {NoStop}%
\bibitem [{\citenamefont {Sathyaprakash}\ \emph {et~al.}(2012)\citenamefont {Sathyaprakash} \emph {et~al.}}]{Sathyaprakash:2012jk}%
  \BibitemOpen
  \bibfield  {author} {\bibinfo {author} {\bibfnamefont {B.}~\bibnamefont {Sathyaprakash}} \emph {et~al.},\ }\bibfield  {title} {\bibinfo {title} {{Scientific Objectives of Einstein Telescope}},\ }\href {https://doi.org/10.1088/0264-9381/29/12/124013} {\bibfield  {journal} {\bibinfo  {journal} {Class. Quant. Grav.}\ }\textbf {\bibinfo {volume} {29}},\ \bibinfo {pages} {124013} (\bibinfo {year} {2012})},\ \bibinfo {note} {[Erratum: Class.Quant.Grav. 30, 079501 (2013)]},\ \Eprint {https://arxiv.org/abs/1206.0331} {arXiv:1206.0331 [gr-qc]} \BibitemShut {NoStop}%
\bibitem [{\citenamefont {Amaro-Seoane}\ \emph {et~al.}(2017)\citenamefont {Amaro-Seoane} \emph {et~al.}}]{LISA:2017pwj}%
  \BibitemOpen
  \bibfield  {author} {\bibinfo {author} {\bibfnamefont {P.}~\bibnamefont {Amaro-Seoane}} \emph {et~al.} (\bibinfo {collaboration} {LISA}),\ }\bibfield  {title} {\bibinfo {title} {{Laser Interferometer Space Antenna}},\ }\href@noop {} {\  (\bibinfo {year} {2017})},\ \Eprint {https://arxiv.org/abs/1702.00786} {arXiv:1702.00786 [astro-ph.IM]} \BibitemShut {NoStop}%
\bibitem [{\citenamefont {Arun}\ \emph {et~al.}(2022)\citenamefont {Arun} \emph {et~al.}}]{LISA:2022kgy}%
  \BibitemOpen
  \bibfield  {author} {\bibinfo {author} {\bibfnamefont {K.~G.}\ \bibnamefont {Arun}} \emph {et~al.} (\bibinfo {collaboration} {LISA}),\ }\bibfield  {title} {\bibinfo {title} {{New horizons for fundamental physics with LISA}},\ }\href {https://doi.org/10.1007/s41114-022-00036-9} {\bibfield  {journal} {\bibinfo  {journal} {Living Rev. Rel.}\ }\textbf {\bibinfo {volume} {25}},\ \bibinfo {pages} {4} (\bibinfo {year} {2022})},\ \Eprint {https://arxiv.org/abs/2205.01597} {arXiv:2205.01597 [gr-qc]} \BibitemShut {NoStop}%
\bibitem [{\citenamefont {Barausse}\ \emph {et~al.}(2020)\citenamefont {Barausse} \emph {et~al.}}]{Barausse:2020rsu}%
  \BibitemOpen
  \bibfield  {author} {\bibinfo {author} {\bibfnamefont {E.}~\bibnamefont {Barausse}} \emph {et~al.},\ }\bibfield  {title} {\bibinfo {title} {{Prospects for Fundamental Physics with LISA}},\ }\href {https://doi.org/10.1007/s10714-020-02691-1} {\bibfield  {journal} {\bibinfo  {journal} {Gen. Rel. Grav.}\ }\textbf {\bibinfo {volume} {52}},\ \bibinfo {pages} {81} (\bibinfo {year} {2020})},\ \Eprint {https://arxiv.org/abs/2001.09793} {arXiv:2001.09793 [gr-qc]} \BibitemShut {NoStop}%
\bibitem [{\citenamefont {Arrechea}\ \emph {et~al.}(2022{\natexlab{a}})\citenamefont {Arrechea}, \citenamefont {Barcel\'o}, \citenamefont {Carballo-Rubio},\ and\ \citenamefont {Garay}}]{Arrecheaetal2022}%
  \BibitemOpen
  \bibfield  {author} {\bibinfo {author} {\bibfnamefont {J.}~\bibnamefont {Arrechea}}, \bibinfo {author} {\bibfnamefont {C.}~\bibnamefont {Barcel\'o}}, \bibinfo {author} {\bibfnamefont {R.}~\bibnamefont {Carballo-Rubio}},\ and\ \bibinfo {author} {\bibfnamefont {L.~J.}\ \bibnamefont {Garay}},\ }\bibfield  {title} {\bibinfo {title} {{Semiclassical relativistic stars}},\ }\href {https://doi.org/10.1038/s41598-022-19836-8} {\bibfield  {journal} {\bibinfo  {journal} {Sci. Rep.}\ }\textbf {\bibinfo {volume} {12}},\ \bibinfo {pages} {15958} (\bibinfo {year} {2022}{\natexlab{a}})},\ \Eprint {https://arxiv.org/abs/2110.15808} {arXiv:2110.15808 [gr-qc]} \BibitemShut {NoStop}%
\bibitem [{\citenamefont {Arrechea}\ \emph {et~al.}(2023)\citenamefont {Arrechea}, \citenamefont {Barcel\'o}, \citenamefont {Carballo-Rubio},\ and\ \citenamefont {Garay}}]{Arrecheaetal2023}%
  \BibitemOpen
  \bibfield  {author} {\bibinfo {author} {\bibfnamefont {J.}~\bibnamefont {Arrechea}}, \bibinfo {author} {\bibfnamefont {C.}~\bibnamefont {Barcel\'o}}, \bibinfo {author} {\bibfnamefont {R.}~\bibnamefont {Carballo-Rubio}},\ and\ \bibinfo {author} {\bibfnamefont {L.~J.}\ \bibnamefont {Garay}},\ }\bibfield  {title} {\bibinfo {title} {{Ultracompact horizonless objects in order-reduced semiclassical gravity}},\ }\href@noop {} {\  (\bibinfo {year} {2023})},\ \Eprint {https://arxiv.org/abs/2310.12668} {arXiv:2310.12668 [gr-qc]} \BibitemShut {NoStop}%
\bibitem [{\citenamefont {Misner}\ and\ \citenamefont {Sharp}(1964)}]{MisnerSharp1964}%
  \BibitemOpen
  \bibfield  {author} {\bibinfo {author} {\bibfnamefont {C.~W.}\ \bibnamefont {Misner}}\ and\ \bibinfo {author} {\bibfnamefont {D.~H.}\ \bibnamefont {Sharp}},\ }\bibfield  {title} {\bibinfo {title} {{Relativistic equations for adiabatic, spherically symmetric gravitational collapse}},\ }\href {https://doi.org/10.1103/PhysRev.136.B571} {\bibfield  {journal} {\bibinfo  {journal} {Phys. Rev.}\ }\textbf {\bibinfo {volume} {136}},\ \bibinfo {pages} {B571} (\bibinfo {year} {1964})}\BibitemShut {NoStop}%
%%CITATION = PHRVA,136,B571;%%
\bibitem [{\citenamefont {{Hernandez}}\ and\ \citenamefont {{Misner}}(1966)}]{HernandezMisner1966}%
  \BibitemOpen
  \bibfield  {author} {\bibinfo {author} {\bibfnamefont {J.}~\bibnamefont {{Hernandez}}, \bibfnamefont {Walter~C.}}\ and\ \bibinfo {author} {\bibfnamefont {C.~W.}\ \bibnamefont {{Misner}}},\ }\bibfield  {title} {\bibinfo {title} {{Observer Time as a Coordinate in Relativistic Spherical Hydrodynamics}},\ }\href {https://doi.org/10.1086/148525} {\bibfield  {journal} {\bibinfo  {journal} {\apj}\ }\textbf {\bibinfo {volume} {143}},\ \bibinfo {pages} {452} (\bibinfo {year} {1966})}\BibitemShut {NoStop}%
\bibitem [{\citenamefont {Hiscock}(1988)}]{Hiscock1988}%
  \BibitemOpen
  \bibfield  {author} {\bibinfo {author} {\bibfnamefont {W.~A.}\ \bibnamefont {Hiscock}},\ }\bibfield  {title} {\bibinfo {title} {Gravitational vacuum polarization around static spherical stars},\ }\href {https://doi.org/10.1103/PhysRevD.37.2142} {\bibfield  {journal} {\bibinfo  {journal} {Phys. Rev. D}\ }\textbf {\bibinfo {volume} {37}},\ \bibinfo {pages} {2142} (\bibinfo {year} {1988})}\BibitemShut {NoStop}%
\bibitem [{\citenamefont {Reyes}\ and\ \citenamefont {Tomaselli}(2023)}]{ReyesTomaselli2023}%
  \BibitemOpen
  \bibfield  {author} {\bibinfo {author} {\bibfnamefont {I.~A.}\ \bibnamefont {Reyes}}\ and\ \bibinfo {author} {\bibfnamefont {G.~M.}\ \bibnamefont {Tomaselli}},\ }\bibfield  {title} {\bibinfo {title} {{Quantum field theory on compact stars near the Buchdahl limit}},\ }\href {https://doi.org/10.1103/PhysRevD.108.065006} {\bibfield  {journal} {\bibinfo  {journal} {Phys. Rev. D}\ }\textbf {\bibinfo {volume} {108}},\ \bibinfo {pages} {065006} (\bibinfo {year} {2023})},\ \Eprint {https://arxiv.org/abs/2301.00826} {arXiv:2301.00826 [gr-qc]} \BibitemShut {NoStop}%
\bibitem [{\citenamefont {Reyes}(2023)}]{Reyes2023}%
  \BibitemOpen
  \bibfield  {author} {\bibinfo {author} {\bibfnamefont {I.~A.}\ \bibnamefont {Reyes}},\ }\bibfield  {title} {\bibinfo {title} {{Trace anomaly and compact stars}},\ }\href@noop {} {\  (\bibinfo {year} {2023})},\ \Eprint {https://arxiv.org/abs/2308.07363} {arXiv:2308.07363 [gr-qc]} \BibitemShut {NoStop}%
\bibitem [{\citenamefont {Boulware}(1975)}]{Boulware1974}%
  \BibitemOpen
  \bibfield  {author} {\bibinfo {author} {\bibfnamefont {D.~G.}\ \bibnamefont {Boulware}},\ }\bibfield  {title} {\bibinfo {title} {Quantum field theory in schwarzschild and rindler spaces},\ }\href {https://doi.org/10.1103/PhysRevD.11.1404} {\bibfield  {journal} {\bibinfo  {journal} {Phys. Rev. D}\ }\textbf {\bibinfo {volume} {11}},\ \bibinfo {pages} {1404} (\bibinfo {year} {1975})}\BibitemShut {NoStop}%
\bibitem [{\citenamefont {Schwarzschild}(1916)}]{Schwarzschild1916b}%
  \BibitemOpen
  \bibfield  {author} {\bibinfo {author} {\bibfnamefont {K.}~\bibnamefont {Schwarzschild}},\ }\bibfield  {title} {\bibinfo {title} {{On the gravitational field of a sphere of incompressible fluid according to Einstein's theory}},\ }\href@noop {} {\bibfield  {journal} {\bibinfo  {journal} {Sitzungsber. Preuss. Akad. Wiss. Berlin (Math. Phys.)}\ }\textbf {\bibinfo {volume} {1916}},\ \bibinfo {pages} {424} (\bibinfo {year} {1916})},\ \Eprint {https://arxiv.org/abs/physics/9912033} {arXiv:physics/9912033 [physics.hist-ph]} \BibitemShut {NoStop}%
%%CITATION = PHYSICS/9912033;%%
\bibitem [{\citenamefont {Cardoso}\ \emph {et~al.}(2014{\natexlab{a}})\citenamefont {Cardoso}, \citenamefont {Crispino}, \citenamefont {Macedo}, \citenamefont {Okawa},\ and\ \citenamefont {Pani}}]{Cardosoetal2014}%
  \BibitemOpen
  \bibfield  {author} {\bibinfo {author} {\bibfnamefont {V.}~\bibnamefont {Cardoso}}, \bibinfo {author} {\bibfnamefont {L.~C.~B.}\ \bibnamefont {Crispino}}, \bibinfo {author} {\bibfnamefont {C.~F.~B.}\ \bibnamefont {Macedo}}, \bibinfo {author} {\bibfnamefont {H.}~\bibnamefont {Okawa}},\ and\ \bibinfo {author} {\bibfnamefont {P.}~\bibnamefont {Pani}},\ }\bibfield  {title} {\bibinfo {title} {Light rings as observational evidence for event horizons: Long-lived modes, ergoregions and nonlinear instabilities of ultracompact objects},\ }\href {https://doi.org/10.1103/PhysRevD.90.044069} {\bibfield  {journal} {\bibinfo  {journal} {Phys. Rev. D}\ }\textbf {\bibinfo {volume} {90}},\ \bibinfo {pages} {044069} (\bibinfo {year} {2014}{\natexlab{a}})}\BibitemShut {NoStop}%
\bibitem [{\citenamefont {Cunha}\ \emph {et~al.}(2017{\natexlab{a}})\citenamefont {Cunha}, \citenamefont {Berti},\ and\ \citenamefont {Herdeiro}}]{Cunhaetal2017}%
  \BibitemOpen
  \bibfield  {author} {\bibinfo {author} {\bibfnamefont {P.~V.~P.}\ \bibnamefont {Cunha}}, \bibinfo {author} {\bibfnamefont {E.}~\bibnamefont {Berti}},\ and\ \bibinfo {author} {\bibfnamefont {C.~A.~R.}\ \bibnamefont {Herdeiro}},\ }\bibfield  {title} {\bibinfo {title} {{Light-Ring Stability for Ultracompact Objects}},\ }\href {https://doi.org/10.1103/PhysRevLett.119.251102} {\bibfield  {journal} {\bibinfo  {journal} {Phys. Rev. Lett.}\ }\textbf {\bibinfo {volume} {119}},\ \bibinfo {pages} {251102} (\bibinfo {year} {2017}{\natexlab{a}})},\ \Eprint {https://arxiv.org/abs/1708.04211} {arXiv:1708.04211 [gr-qc]} \BibitemShut {NoStop}%
\bibitem [{\citenamefont {Guo}\ \emph {et~al.}(2022{\natexlab{a}})\citenamefont {Guo}, \citenamefont {Wang}, \citenamefont {Wu},\ and\ \citenamefont {Yang}}]{Guoetal2022}%
  \BibitemOpen
  \bibfield  {author} {\bibinfo {author} {\bibfnamefont {G.}~\bibnamefont {Guo}}, \bibinfo {author} {\bibfnamefont {P.}~\bibnamefont {Wang}}, \bibinfo {author} {\bibfnamefont {H.}~\bibnamefont {Wu}},\ and\ \bibinfo {author} {\bibfnamefont {H.}~\bibnamefont {Yang}},\ }\bibfield  {title} {\bibinfo {title} {{Echoes from hairy black holes}},\ }\href {https://doi.org/10.1007/JHEP06(2022)073} {\bibfield  {journal} {\bibinfo  {journal} {JHEP}\ }\textbf {\bibinfo {volume} {06}},\ \bibinfo {pages} {073}},\ \Eprint {https://arxiv.org/abs/2204.00982} {arXiv:2204.00982 [gr-qc]} \BibitemShut {NoStop}%
\bibitem [{\citenamefont {Abedi}\ \emph {et~al.}(2017{\natexlab{b}})\citenamefont {Abedi}, \citenamefont {Dykaar},\ and\ \citenamefont {Afshordi}}]{AbediAfshordi2016}%
  \BibitemOpen
  \bibfield  {author} {\bibinfo {author} {\bibfnamefont {J.}~\bibnamefont {Abedi}}, \bibinfo {author} {\bibfnamefont {H.}~\bibnamefont {Dykaar}},\ and\ \bibinfo {author} {\bibfnamefont {N.}~\bibnamefont {Afshordi}},\ }\bibfield  {title} {\bibinfo {title} {{Echoes from the Abyss: Tentative evidence for Planck-scale structure at black hole horizons}},\ }\href {https://doi.org/10.1103/PhysRevD.96.082004} {\bibfield  {journal} {\bibinfo  {journal} {Phys. Rev. D}\ }\textbf {\bibinfo {volume} {96}},\ \bibinfo {pages} {082004} (\bibinfo {year} {2017}{\natexlab{b}})},\ \Eprint {https://arxiv.org/abs/1612.00266} {arXiv:1612.00266 [gr-qc]} \BibitemShut {NoStop}%
\bibitem [{\citenamefont {Zimmerman}\ \emph {et~al.}(2023)\citenamefont {Zimmerman}, \citenamefont {George},\ and\ \citenamefont {Chen}}]{Zimmermanetal2023}%
  \BibitemOpen
  \bibfield  {author} {\bibinfo {author} {\bibfnamefont {A.}~\bibnamefont {Zimmerman}}, \bibinfo {author} {\bibfnamefont {R.~N.}\ \bibnamefont {George}},\ and\ \bibinfo {author} {\bibfnamefont {Y.}~\bibnamefont {Chen}},\ }\bibfield  {title} {\bibinfo {title} {{Rogue echoes from exotic compact objects}},\ }\href@noop {} {\  (\bibinfo {year} {2023})},\ \Eprint {https://arxiv.org/abs/2306.11166} {arXiv:2306.11166 [gr-qc]} \BibitemShut {NoStop}%
\bibitem [{\citenamefont {Karageorgis}\ and\ \citenamefont {Stalker}(2008)}]{Karageorgis2007}%
  \BibitemOpen
  \bibfield  {author} {\bibinfo {author} {\bibfnamefont {P.}~\bibnamefont {Karageorgis}}\ and\ \bibinfo {author} {\bibfnamefont {J.~G.}\ \bibnamefont {Stalker}},\ }\bibfield  {title} {\bibinfo {title} {{Sharp bounds on 2m/r for static spherical objects}},\ }\href {https://doi.org/10.1088/0264-9381/25/19/195021} {\bibfield  {journal} {\bibinfo  {journal} {Class. Quant. Grav.}\ }\textbf {\bibinfo {volume} {25}},\ \bibinfo {pages} {195021} (\bibinfo {year} {2008})},\ \Eprint {https://arxiv.org/abs/0707.3632} {arXiv:0707.3632 [gr-qc]} \BibitemShut {NoStop}%
%%CITATION = ARXIV:0707.3632;%%
\bibitem [{\citenamefont {Andreasson}(2008)}]{Andreasson2008}%
  \BibitemOpen
  \bibfield  {author} {\bibinfo {author} {\bibfnamefont {H.}~\bibnamefont {Andreasson}},\ }\bibfield  {title} {\bibinfo {title} {{Sharp bounds on 2m/r of general spherically symmetric static objects}},\ }\href {https://doi.org/10.1016/j.jde.2008.05.010} {\bibfield  {journal} {\bibinfo  {journal} {J. Diff. Eq.}\ }\textbf {\bibinfo {volume} {245}},\ \bibinfo {pages} {2243} (\bibinfo {year} {2008})},\ \Eprint {https://arxiv.org/abs/gr-qc/0702137} {arXiv:gr-qc/0702137} \BibitemShut {NoStop}%
\bibitem [{\citenamefont {Urbano}\ and\ \citenamefont {Veerm\"ae}(2019)}]{UrbanoVeermae2018}%
  \BibitemOpen
  \bibfield  {author} {\bibinfo {author} {\bibfnamefont {A.}~\bibnamefont {Urbano}}\ and\ \bibinfo {author} {\bibfnamefont {H.}~\bibnamefont {Veerm\"ae}},\ }\bibfield  {title} {\bibinfo {title} {{On gravitational echoes from ultracompact exotic stars}},\ }\href {https://doi.org/10.1088/1475-7516/2019/04/011} {\bibfield  {journal} {\bibinfo  {journal} {JCAP}\ }\textbf {\bibinfo {volume} {04}},\ \bibinfo {pages} {011}},\ \Eprint {https://arxiv.org/abs/1810.07137} {arXiv:1810.07137 [gr-qc]} \BibitemShut {NoStop}%
\bibitem [{\citenamefont {Raposo}\ \emph {et~al.}(2019{\natexlab{b}})\citenamefont {Raposo}, \citenamefont {Pani}, \citenamefont {Bezares}, \citenamefont {Palenzuela},\ and\ \citenamefont {Cardoso}}]{Raposoetal2019}%
  \BibitemOpen
  \bibfield  {author} {\bibinfo {author} {\bibfnamefont {G.}~\bibnamefont {Raposo}}, \bibinfo {author} {\bibfnamefont {P.}~\bibnamefont {Pani}}, \bibinfo {author} {\bibfnamefont {M.}~\bibnamefont {Bezares}}, \bibinfo {author} {\bibfnamefont {C.}~\bibnamefont {Palenzuela}},\ and\ \bibinfo {author} {\bibfnamefont {V.}~\bibnamefont {Cardoso}},\ }\bibfield  {title} {\bibinfo {title} {Anisotropic stars as ultracompact objects in general relativity},\ }\href {https://doi.org/10.1103/PhysRevD.99.104072} {\bibfield  {journal} {\bibinfo  {journal} {Phys. Rev. D}\ }\textbf {\bibinfo {volume} {99}},\ \bibinfo {pages} {104072} (\bibinfo {year} {2019}{\natexlab{b}})}\BibitemShut {NoStop}%
\bibitem [{\citenamefont {Howard}\ and\ \citenamefont {Candelas}(1984)}]{CandelasHoward1984}%
  \BibitemOpen
  \bibfield  {author} {\bibinfo {author} {\bibfnamefont {K.~W.}\ \bibnamefont {Howard}}\ and\ \bibinfo {author} {\bibfnamefont {P.}~\bibnamefont {Candelas}},\ }\bibfield  {title} {\bibinfo {title} {Quantum stress tensor in schwarzschild space-time},\ }\href {https://doi.org/10.1103/PhysRevLett.53.403} {\bibfield  {journal} {\bibinfo  {journal} {Phys. Rev. Lett.}\ }\textbf {\bibinfo {volume} {53}},\ \bibinfo {pages} {403} (\bibinfo {year} {1984})}\BibitemShut {NoStop}%
\bibitem [{\citenamefont {Anderson}\ \emph {et~al.}(1995)\citenamefont {Anderson}, \citenamefont {Hiscock},\ and\ \citenamefont {Samuel}}]{Andersonetal1995}%
  \BibitemOpen
  \bibfield  {author} {\bibinfo {author} {\bibfnamefont {P.~R.}\ \bibnamefont {Anderson}}, \bibinfo {author} {\bibfnamefont {W.~A.}\ \bibnamefont {Hiscock}},\ and\ \bibinfo {author} {\bibfnamefont {D.~A.}\ \bibnamefont {Samuel}},\ }\bibfield  {title} {\bibinfo {title} {Stress-energy tensor of quantized scalar fields in static spherically symmetric spacetimes},\ }\href {https://doi.org/10.1103/PhysRevD.51.4337} {\bibfield  {journal} {\bibinfo  {journal} {Phys. Rev. D}\ }\textbf {\bibinfo {volume} {51}},\ \bibinfo {pages} {4337} (\bibinfo {year} {1995})}\BibitemShut {NoStop}%
\bibitem [{\citenamefont {Levi}\ and\ \citenamefont {Ori}(2016)}]{LeviOri2016}%
  \BibitemOpen
  \bibfield  {author} {\bibinfo {author} {\bibfnamefont {A.}~\bibnamefont {Levi}}\ and\ \bibinfo {author} {\bibfnamefont {A.}~\bibnamefont {Ori}},\ }\bibfield  {title} {\bibinfo {title} {Versatile method for renormalized stress-energy computation in black-hole spacetimes},\ }\href {https://doi.org/10.1103/PhysRevLett.117.231101} {\bibfield  {journal} {\bibinfo  {journal} {Phys. Rev. Lett.}\ }\textbf {\bibinfo {volume} {117}},\ \bibinfo {pages} {231101} (\bibinfo {year} {2016})}\BibitemShut {NoStop}%
\bibitem [{\citenamefont {Taylor}\ \emph {et~al.}(2022)\citenamefont {Taylor}, \citenamefont {Breen},\ and\ \citenamefont {Ottewill}}]{Taylor:2022sly}%
  \BibitemOpen
  \bibfield  {author} {\bibinfo {author} {\bibfnamefont {P.}~\bibnamefont {Taylor}}, \bibinfo {author} {\bibfnamefont {C.}~\bibnamefont {Breen}},\ and\ \bibinfo {author} {\bibfnamefont {A.}~\bibnamefont {Ottewill}},\ }\bibfield  {title} {\bibinfo {title} {{Mode-sum prescription for the renormalized stress energy tensor on black hole spacetimes}},\ }\href {https://doi.org/10.1103/PhysRevD.106.065023} {\bibfield  {journal} {\bibinfo  {journal} {Phys. Rev. D}\ }\textbf {\bibinfo {volume} {106}},\ \bibinfo {pages} {065023} (\bibinfo {year} {2022})},\ \Eprint {https://arxiv.org/abs/2201.05174} {arXiv:2201.05174 [gr-qc]} \BibitemShut {NoStop}%
\bibitem [{\citenamefont {Flanagan}\ and\ \citenamefont {Wald}(1996)}]{FlanaganWald1996}%
  \BibitemOpen
  \bibfield  {author} {\bibinfo {author} {\bibfnamefont {E.~E.}\ \bibnamefont {Flanagan}}\ and\ \bibinfo {author} {\bibfnamefont {R.~M.}\ \bibnamefont {Wald}},\ }\bibfield  {title} {\bibinfo {title} {{Does back reaction enforce the averaged null energy condition in semiclassical gravity?}},\ }\href {https://doi.org/10.1103/PhysRevD.54.6233} {\bibfield  {journal} {\bibinfo  {journal} {Phys. Rev. D}\ }\textbf {\bibinfo {volume} {54}},\ \bibinfo {pages} {6233} (\bibinfo {year} {1996})},\ \Eprint {https://arxiv.org/abs/gr-qc/9602052} {arXiv:gr-qc/9602052} \BibitemShut {NoStop}%
\bibitem [{\citenamefont {{Thorne}}\ and\ \citenamefont {{Campolattaro}}(1967)}]{ThorneCampolattaro1967}%
  \BibitemOpen
  \bibfield  {author} {\bibinfo {author} {\bibfnamefont {K.~S.}\ \bibnamefont {{Thorne}}}\ and\ \bibinfo {author} {\bibfnamefont {A.}~\bibnamefont {{Campolattaro}}},\ }\href {https://doi.org/10.1086/149288} {\bibinfo {title} {{Non-Radial Pulsation of General-Relativistic Stellar Models. I. Analytic Analysis for L $\geq$ 2}}} (\bibinfo {year} {1967})\BibitemShut {NoStop}%
\bibitem [{\citenamefont {Chandrasekhar}\ and\ \citenamefont {Ferrari}(1991)}]{ChandrasekharFerrari1991}%
  \BibitemOpen
  \bibfield  {author} {\bibinfo {author} {\bibfnamefont {S.}~\bibnamefont {Chandrasekhar}}\ and\ \bibinfo {author} {\bibfnamefont {V.}~\bibnamefont {Ferrari}},\ }\bibfield  {title} {\bibinfo {title} {{On the non-radial oscillations of a star}},\ }\href {https://doi.org/10.1098/rspa.1991.0016} {\bibfield  {journal} {\bibinfo  {journal} {Proc. Roy. Soc. Lond. A}\ }\textbf {\bibinfo {volume} {432}},\ \bibinfo {pages} {247} (\bibinfo {year} {1991})}\BibitemShut {NoStop}%
\bibitem [{\citenamefont {Nollert}(1999)}]{Nollert1999}%
  \BibitemOpen
  \bibfield  {author} {\bibinfo {author} {\bibfnamefont {H.-P.}\ \bibnamefont {Nollert}},\ }\bibfield  {title} {\bibinfo {title} {Quasinormal modes: the characteristic `sound' of black holes and neutron stars},\ }\href {https://api.semanticscholar.org/CorpusID:123577194} {\bibfield  {journal} {\bibinfo  {journal} {Classical and Quantum Gravity}\ }\textbf {\bibinfo {volume} {16}},\ \bibinfo {pages} {R159 } (\bibinfo {year} {1999})}\BibitemShut {NoStop}%
\bibitem [{\citenamefont {Karlovini}(2002)}]{Karlovini:2001fc}%
  \BibitemOpen
  \bibfield  {author} {\bibinfo {author} {\bibfnamefont {M.}~\bibnamefont {Karlovini}},\ }\bibfield  {title} {\bibinfo {title} {{Axial perturbations of general spherically symmetric space-times}},\ }\href {https://doi.org/10.1088/0264-9381/19/8/305} {\bibfield  {journal} {\bibinfo  {journal} {Class. Quant. Grav.}\ }\textbf {\bibinfo {volume} {19}},\ \bibinfo {pages} {2125} (\bibinfo {year} {2002})},\ \Eprint {https://arxiv.org/abs/gr-qc/0111066} {arXiv:gr-qc/0111066} \BibitemShut {NoStop}%
\bibitem [{\citenamefont {Medved}\ \emph {et~al.}(2004)\citenamefont {Medved}, \citenamefont {Martin},\ and\ \citenamefont {Visser}}]{Medved:2003pr}%
  \BibitemOpen
  \bibfield  {author} {\bibinfo {author} {\bibfnamefont {A.~J.~M.}\ \bibnamefont {Medved}}, \bibinfo {author} {\bibfnamefont {D.}~\bibnamefont {Martin}},\ and\ \bibinfo {author} {\bibfnamefont {M.}~\bibnamefont {Visser}},\ }\bibfield  {title} {\bibinfo {title} {{Dirty black holes: Quasinormal modes for `squeezed' horizons}},\ }\href {https://doi.org/10.1088/0264-9381/21/9/013} {\bibfield  {journal} {\bibinfo  {journal} {Class. Quantum Grav.}\ }\textbf {\bibinfo {volume} {21}},\ \bibinfo {pages} {2393} (\bibinfo {year} {2004})},\ \Eprint {https://arxiv.org/abs/gr-qc/0310097} {arXiv:gr-qc/0310097} \BibitemShut {NoStop}%
\bibitem [{\citenamefont {Cardoso}\ and\ \citenamefont {Pani}(2017{\natexlab{b}})}]{Cardoso:2017cqb}%
  \BibitemOpen
  \bibfield  {author} {\bibinfo {author} {\bibfnamefont {V.}~\bibnamefont {Cardoso}}\ and\ \bibinfo {author} {\bibfnamefont {P.}~\bibnamefont {Pani}},\ }\bibfield  {title} {\bibinfo {title} {{Tests for the existence of black holes through gravitational wave echoes}},\ }\href {https://doi.org/10.1038/s41550-017-0225-y} {\bibfield  {journal} {\bibinfo  {journal} {Nature Astron.}\ }\textbf {\bibinfo {volume} {1}},\ \bibinfo {pages} {586} (\bibinfo {year} {2017}{\natexlab{b}})},\ \Eprint {https://arxiv.org/abs/1709.01525} {arXiv:1709.01525 [gr-qc]} \BibitemShut {NoStop}%
\bibitem [{\citenamefont {Annulli}\ \emph {et~al.}(2022)\citenamefont {Annulli}, \citenamefont {Cardoso},\ and\ \citenamefont {Gualtieri}}]{Annulli:2021ccn}%
  \BibitemOpen
  \bibfield  {author} {\bibinfo {author} {\bibfnamefont {L.}~\bibnamefont {Annulli}}, \bibinfo {author} {\bibfnamefont {V.}~\bibnamefont {Cardoso}},\ and\ \bibinfo {author} {\bibfnamefont {L.}~\bibnamefont {Gualtieri}},\ }\bibfield  {title} {\bibinfo {title} {{Applications of the close-limit approximation: horizonless compact objects and scalar fields}},\ }\href {https://doi.org/10.1088/1361-6382/ac6410} {\bibfield  {journal} {\bibinfo  {journal} {Class. Quant. Grav.}\ }\textbf {\bibinfo {volume} {39}},\ \bibinfo {pages} {105005} (\bibinfo {year} {2022})},\ \Eprint {https://arxiv.org/abs/2104.11236} {arXiv:2104.11236 [gr-qc]} \BibitemShut {NoStop}%
\bibitem [{\citenamefont {Chen}\ and\ \citenamefont {Yokokura}(2024)}]{Chen:2024ibc}%
  \BibitemOpen
  \bibfield  {author} {\bibinfo {author} {\bibfnamefont {C.-Y.}\ \bibnamefont {Chen}}\ and\ \bibinfo {author} {\bibfnamefont {Y.}~\bibnamefont {Yokokura}},\ }\bibfield  {title} {\bibinfo {title} {{Imaging a semiclassical horizonless compact object with strong redshift}},\ }\href {https://doi.org/10.1103/PhysRevD.109.104058} {\bibfield  {journal} {\bibinfo  {journal} {Phys. Rev. D}\ }\textbf {\bibinfo {volume} {109}},\ \bibinfo {pages} {104058} (\bibinfo {year} {2024})},\ \Eprint {https://arxiv.org/abs/2403.09388} {arXiv:2403.09388 [gr-qc]} \BibitemShut {NoStop}%
\bibitem [{\citenamefont {Kokkotas}\ and\ \citenamefont {Schmidt}(1999)}]{KokkotasSchmidt1999}%
  \BibitemOpen
  \bibfield  {author} {\bibinfo {author} {\bibfnamefont {K.~D.}\ \bibnamefont {Kokkotas}}\ and\ \bibinfo {author} {\bibfnamefont {B.~G.}\ \bibnamefont {Schmidt}},\ }\bibfield  {title} {\bibinfo {title} {{Quasinormal modes of stars and black holes}},\ }\href {https://doi.org/10.12942/lrr-1999-2} {\bibfield  {journal} {\bibinfo  {journal} {Living Rev. Rel.}\ }\textbf {\bibinfo {volume} {2}},\ \bibinfo {pages} {2} (\bibinfo {year} {1999})},\ \Eprint {https://arxiv.org/abs/gr-qc/9909058} {arXiv:gr-qc/9909058} \BibitemShut {NoStop}%
\bibitem [{\citenamefont {Konoplya}\ \emph {et~al.}(2019)\citenamefont {Konoplya}, \citenamefont {Posada}, \citenamefont {Stuchl\'\i{}k},\ and\ \citenamefont {Zhidenko}}]{Konoplyaetal2019}%
  \BibitemOpen
  \bibfield  {author} {\bibinfo {author} {\bibfnamefont {R.~A.}\ \bibnamefont {Konoplya}}, \bibinfo {author} {\bibfnamefont {C.}~\bibnamefont {Posada}}, \bibinfo {author} {\bibfnamefont {Z.}~\bibnamefont {Stuchl\'\i{}k}},\ and\ \bibinfo {author} {\bibfnamefont {A.}~\bibnamefont {Zhidenko}},\ }\bibfield  {title} {\bibinfo {title} {{Stable Schwarzschild stars as black-hole mimickers}},\ }\href {https://doi.org/10.1103/PhysRevD.100.044027} {\bibfield  {journal} {\bibinfo  {journal} {Phys. Rev. D}\ }\textbf {\bibinfo {volume} {100}},\ \bibinfo {pages} {044027} (\bibinfo {year} {2019})},\ \Eprint {https://arxiv.org/abs/1905.08097} {arXiv:1905.08097 [gr-qc]} \BibitemShut {NoStop}%
\bibitem [{\citenamefont {{Chandrasekhar}}\ and\ \citenamefont {{Detweiler}}(1975)}]{ChandrasekharDetweiler1975}%
  \BibitemOpen
  \bibfield  {author} {\bibinfo {author} {\bibfnamefont {S.}~\bibnamefont {{Chandrasekhar}}}\ and\ \bibinfo {author} {\bibfnamefont {S.}~\bibnamefont {{Detweiler}}},\ }\bibfield  {title} {\bibinfo {title} {{The Quasi-Normal Modes of the Schwarzschild Black Hole}},\ }\href {https://doi.org/10.1098/rspa.1975.0112} {\bibfield  {journal} {\bibinfo  {journal} {Proceedings of the Royal Society of London Series A}\ }\textbf {\bibinfo {volume} {344}},\ \bibinfo {pages} {441} (\bibinfo {year} {1975})}\BibitemShut {NoStop}%
\bibitem [{\citenamefont {Molina}\ \emph {et~al.}(2010)\citenamefont {Molina}, \citenamefont {Pani}, \citenamefont {Cardoso},\ and\ \citenamefont {Gualtieri}}]{Molinaetal2010}%
  \BibitemOpen
  \bibfield  {author} {\bibinfo {author} {\bibfnamefont {C.}~\bibnamefont {Molina}}, \bibinfo {author} {\bibfnamefont {P.}~\bibnamefont {Pani}}, \bibinfo {author} {\bibfnamefont {V.}~\bibnamefont {Cardoso}},\ and\ \bibinfo {author} {\bibfnamefont {L.}~\bibnamefont {Gualtieri}},\ }\bibfield  {title} {\bibinfo {title} {{Gravitational signature of Schwarzschild black holes in dynamical Chern-Simons gravity}},\ }\href {https://doi.org/10.1103/PhysRevD.81.124021} {\bibfield  {journal} {\bibinfo  {journal} {Phys. Rev. D}\ }\textbf {\bibinfo {volume} {81}},\ \bibinfo {pages} {124021} (\bibinfo {year} {2010})},\ \Eprint {https://arxiv.org/abs/1004.4007} {arXiv:1004.4007 [gr-qc]} \BibitemShut {NoStop}%
\bibitem [{\citenamefont {Arrechea}\ \emph {et~al.}(2022{\natexlab{b}})\citenamefont {Arrechea}, \citenamefont {Barcel\'o}, \citenamefont {Carballo-Rubio},\ and\ \citenamefont {Garay}}]{Arrecheaetal2022b}%
  \BibitemOpen
  \bibfield  {author} {\bibinfo {author} {\bibfnamefont {J.}~\bibnamefont {Arrechea}}, \bibinfo {author} {\bibfnamefont {C.}~\bibnamefont {Barcel\'o}}, \bibinfo {author} {\bibfnamefont {R.}~\bibnamefont {Carballo-Rubio}},\ and\ \bibinfo {author} {\bibfnamefont {L.~J.}\ \bibnamefont {Garay}},\ }\bibfield  {title} {\bibinfo {title} {{Asymptotically flat vacuum solutions in order-reduced semiclassical gravity}},\ }\href@noop {} {\  (\bibinfo {year} {2022}{\natexlab{b}})},\ \Eprint {https://arxiv.org/abs/2212.09375} {arXiv:2212.09375 [gr-qc]} \BibitemShut {NoStop}%
\bibitem [{\citenamefont {Cardoso}\ and\ \citenamefont {Pani}(2019)}]{CardosoPani2019}%
  \BibitemOpen
  \bibfield  {author} {\bibinfo {author} {\bibfnamefont {V.}~\bibnamefont {Cardoso}}\ and\ \bibinfo {author} {\bibfnamefont {P.}~\bibnamefont {Pani}},\ }\bibfield  {title} {\bibinfo {title} {{Testing the nature of dark compact objects: a status report}},\ }\href {https://doi.org/10.1007/s41114-019-0020-4} {\bibfield  {journal} {\bibinfo  {journal} {Living Rev. Rel.}\ }\textbf {\bibinfo {volume} {22}},\ \bibinfo {pages} {4} (\bibinfo {year} {2019})},\ \Eprint {https://arxiv.org/abs/1904.05363} {arXiv:1904.05363 [gr-qc]} \BibitemShut {NoStop}%
%%CITATION = ARXIV:1904.05363;%%
\bibitem [{\citenamefont {Maggio}\ \emph {et~al.}(2021)\citenamefont {Maggio}, \citenamefont {van~de Meent},\ and\ \citenamefont {Pani}}]{Maggioetal2021}%
  \BibitemOpen
  \bibfield  {author} {\bibinfo {author} {\bibfnamefont {E.}~\bibnamefont {Maggio}}, \bibinfo {author} {\bibfnamefont {M.}~\bibnamefont {van~de Meent}},\ and\ \bibinfo {author} {\bibfnamefont {P.}~\bibnamefont {Pani}},\ }\bibfield  {title} {\bibinfo {title} {Extreme mass-ratio inspirals around a spinning horizonless compact object},\ }\href {https://doi.org/10.1103/PhysRevD.104.104026} {\bibfield  {journal} {\bibinfo  {journal} {Phys. Rev. D}\ }\textbf {\bibinfo {volume} {104}},\ \bibinfo {pages} {104026} (\bibinfo {year} {2021})}\BibitemShut {NoStop}%
\bibitem [{\citenamefont {Maggio}(2023)}]{Maggio2023}%
  \BibitemOpen
  \bibfield  {author} {\bibinfo {author} {\bibfnamefont {E.}~\bibnamefont {Maggio}},\ }\bibfield  {title} {\bibinfo {title} {{Probing the Horizon of Black Holes with Gravitational Waves}},\ }\href {https://doi.org/10.1007/978-3-031-31520-6_9} {\bibfield  {journal} {\bibinfo  {journal} {Lect. Notes Phys.}\ }\textbf {\bibinfo {volume} {1017}},\ \bibinfo {pages} {333} (\bibinfo {year} {2023})},\ \Eprint {https://arxiv.org/abs/2310.07368} {arXiv:2310.07368 [gr-qc]} \BibitemShut {NoStop}%
\bibitem [{\citenamefont {Dymnikova}(1992)}]{Dymnikova:1992ux}%
  \BibitemOpen
  \bibfield  {author} {\bibinfo {author} {\bibfnamefont {I.}~\bibnamefont {Dymnikova}},\ }\bibfield  {title} {\bibinfo {title} {{Vacuum nonsingular black hole}},\ }\href {https://doi.org/10.1007/BF00760226} {\bibfield  {journal} {\bibinfo  {journal} {Gen. Rel. Grav.}\ }\textbf {\bibinfo {volume} {24}},\ \bibinfo {pages} {235} (\bibinfo {year} {1992})}\BibitemShut {NoStop}%
\bibitem [{\citenamefont {Dymnikova}(2002)}]{Dymnikova:2001fb}%
  \BibitemOpen
  \bibfield  {author} {\bibinfo {author} {\bibfnamefont {I.}~\bibnamefont {Dymnikova}},\ }\bibfield  {title} {\bibinfo {title} {{Cosmological term as a source of mass}},\ }\href {https://doi.org/10.1088/0264-9381/19/4/306} {\bibfield  {journal} {\bibinfo  {journal} {Class. Quant. Grav.}\ }\textbf {\bibinfo {volume} {19}},\ \bibinfo {pages} {725} (\bibinfo {year} {2002})},\ \Eprint {https://arxiv.org/abs/gr-qc/0112052} {arXiv:gr-qc/0112052} \BibitemShut {NoStop}%
\bibitem [{\citenamefont {Cunha}\ \emph {et~al.}(2017{\natexlab{b}})\citenamefont {Cunha}, \citenamefont {Berti},\ and\ \citenamefont {Herdeiro}}]{Cunha:2017qtt}%
  \BibitemOpen
  \bibfield  {author} {\bibinfo {author} {\bibfnamefont {P.~V.~P.}\ \bibnamefont {Cunha}}, \bibinfo {author} {\bibfnamefont {E.}~\bibnamefont {Berti}},\ and\ \bibinfo {author} {\bibfnamefont {C.~A.~R.}\ \bibnamefont {Herdeiro}},\ }\bibfield  {title} {\bibinfo {title} {{Light-Ring Stability for Ultracompact Objects}},\ }\href {https://doi.org/10.1103/PhysRevLett.119.251102} {\bibfield  {journal} {\bibinfo  {journal} {Phys. Rev. Lett.}\ }\textbf {\bibinfo {volume} {119}},\ \bibinfo {pages} {251102} (\bibinfo {year} {2017}{\natexlab{b}})},\ \Eprint {https://arxiv.org/abs/1708.04211} {arXiv:1708.04211 [gr-qc]} \BibitemShut {NoStop}%
\bibitem [{\citenamefont {Cardoso}\ \emph {et~al.}(2014{\natexlab{b}})\citenamefont {Cardoso}, \citenamefont {Crispino}, \citenamefont {Macedo}, \citenamefont {Okawa},\ and\ \citenamefont {Pani}}]{Cardoso:2014sna}%
  \BibitemOpen
  \bibfield  {author} {\bibinfo {author} {\bibfnamefont {V.}~\bibnamefont {Cardoso}}, \bibinfo {author} {\bibfnamefont {L.~C.~B.}\ \bibnamefont {Crispino}}, \bibinfo {author} {\bibfnamefont {C.~F.~B.}\ \bibnamefont {Macedo}}, \bibinfo {author} {\bibfnamefont {H.}~\bibnamefont {Okawa}},\ and\ \bibinfo {author} {\bibfnamefont {P.}~\bibnamefont {Pani}},\ }\bibfield  {title} {\bibinfo {title} {{Light rings as observational evidence for event horizons: long-lived modes, ergoregions and nonlinear instabilities of ultracompact objects}},\ }\href {https://doi.org/10.1103/PhysRevD.90.044069} {\bibfield  {journal} {\bibinfo  {journal} {Phys. Rev. D}\ }\textbf {\bibinfo {volume} {90}},\ \bibinfo {pages} {044069} (\bibinfo {year} {2014}{\natexlab{b}})},\ \Eprint {https://arxiv.org/abs/1406.5510} {arXiv:1406.5510 [gr-qc]} \BibitemShut {NoStop}%
\bibitem [{\citenamefont {Cunha}\ \emph {et~al.}(2023)\citenamefont {Cunha}, \citenamefont {Herdeiro}, \citenamefont {Radu},\ and\ \citenamefont {Sanchis-Gual}}]{Cunha:2022gde}%
  \BibitemOpen
  \bibfield  {author} {\bibinfo {author} {\bibfnamefont {P.~V.~P.}\ \bibnamefont {Cunha}}, \bibinfo {author} {\bibfnamefont {C.}~\bibnamefont {Herdeiro}}, \bibinfo {author} {\bibfnamefont {E.}~\bibnamefont {Radu}},\ and\ \bibinfo {author} {\bibfnamefont {N.}~\bibnamefont {Sanchis-Gual}},\ }\bibfield  {title} {\bibinfo {title} {{Exotic Compact Objects and the Fate of the Light-Ring Instability}},\ }\href {https://doi.org/10.1103/PhysRevLett.130.061401} {\bibfield  {journal} {\bibinfo  {journal} {Phys. Rev. Lett.}\ }\textbf {\bibinfo {volume} {130}},\ \bibinfo {pages} {061401} (\bibinfo {year} {2023})},\ \Eprint {https://arxiv.org/abs/2207.13713} {arXiv:2207.13713 [gr-qc]} \BibitemShut {NoStop}%
\bibitem [{\citenamefont {Tominaga}\ \emph {et~al.}(1999)\citenamefont {Tominaga}, \citenamefont {Saijo},\ and\ \citenamefont {Maeda}}]{Tominaga:1999iy}%
  \BibitemOpen
  \bibfield  {author} {\bibinfo {author} {\bibfnamefont {K.}~\bibnamefont {Tominaga}}, \bibinfo {author} {\bibfnamefont {M.}~\bibnamefont {Saijo}},\ and\ \bibinfo {author} {\bibfnamefont {K.-i.}\ \bibnamefont {Maeda}},\ }\bibfield  {title} {\bibinfo {title} {{Gravitational waves from a test particle scattered by a neutron star: Axial mode case}},\ }\href {https://doi.org/10.1103/PhysRevD.60.024004} {\bibfield  {journal} {\bibinfo  {journal} {Phys. Rev. D}\ }\textbf {\bibinfo {volume} {60}},\ \bibinfo {pages} {024004} (\bibinfo {year} {1999})},\ \Eprint {https://arxiv.org/abs/gr-qc/9901040} {arXiv:gr-qc/9901040} \BibitemShut {NoStop}%
\bibitem [{\citenamefont {Kokkotas}(1995)}]{Kokkotas:1995av}%
  \BibitemOpen
  \bibfield  {author} {\bibinfo {author} {\bibfnamefont {K.~D.}\ \bibnamefont {Kokkotas}},\ }\bibfield  {title} {\bibinfo {title} {{Pulsating relativistic stars}},\ }in\ \href@noop {} {\emph {\bibinfo {booktitle} {{Les Houches School of Physics: Astrophysical Sources of Gravitational Radiation}}}}\ (\bibinfo {year} {1995})\ pp.\ \bibinfo {pages} {89--102},\ \Eprint {https://arxiv.org/abs/gr-qc/9603024} {arXiv:gr-qc/9603024} \BibitemShut {NoStop}%
\bibitem [{\citenamefont {Ferrari}\ and\ \citenamefont {Kokkotas}(2000)}]{Ferrari:2000sr}%
  \BibitemOpen
  \bibfield  {author} {\bibinfo {author} {\bibfnamefont {V.}~\bibnamefont {Ferrari}}\ and\ \bibinfo {author} {\bibfnamefont {K.~D.}\ \bibnamefont {Kokkotas}},\ }\bibfield  {title} {\bibinfo {title} {{Scattering of particles by neutron stars: Time evolutions for axial perturbations}},\ }\href {https://doi.org/10.1103/PhysRevD.62.107504} {\bibfield  {journal} {\bibinfo  {journal} {Phys. Rev. D}\ }\textbf {\bibinfo {volume} {62}},\ \bibinfo {pages} {107504} (\bibinfo {year} {2000})},\ \Eprint {https://arxiv.org/abs/gr-qc/0008057} {arXiv:gr-qc/0008057} \BibitemShut {NoStop}%
\bibitem [{\citenamefont {Guo}\ \emph {et~al.}(2022{\natexlab{b}})\citenamefont {Guo}, \citenamefont {Wang}, \citenamefont {Wu},\ and\ \citenamefont {Yang}}]{Guo:2022umh}%
  \BibitemOpen
  \bibfield  {author} {\bibinfo {author} {\bibfnamefont {G.}~\bibnamefont {Guo}}, \bibinfo {author} {\bibfnamefont {P.}~\bibnamefont {Wang}}, \bibinfo {author} {\bibfnamefont {H.}~\bibnamefont {Wu}},\ and\ \bibinfo {author} {\bibfnamefont {H.}~\bibnamefont {Yang}},\ }\bibfield  {title} {\bibinfo {title} {{Echoes from hairy black holes}},\ }\href {https://doi.org/10.1007/JHEP06(2022)073} {\bibfield  {journal} {\bibinfo  {journal} {JHEP}\ }\textbf {\bibinfo {volume} {06}},\ \bibinfo {pages} {073}},\ \Eprint {https://arxiv.org/abs/2204.00982} {arXiv:2204.00982 [gr-qc]} \BibitemShut {NoStop}%
\bibitem [{\citenamefont {Chen}\ \emph {et~al.}(2019)\citenamefont {Chen}, \citenamefont {Chen}, \citenamefont {Ma}, \citenamefont {Lo},\ and\ \citenamefont {Sun}}]{Chen:2019hfg}%
  \BibitemOpen
  \bibfield  {author} {\bibinfo {author} {\bibfnamefont {B.}~\bibnamefont {Chen}}, \bibinfo {author} {\bibfnamefont {Y.}~\bibnamefont {Chen}}, \bibinfo {author} {\bibfnamefont {Y.}~\bibnamefont {Ma}}, \bibinfo {author} {\bibfnamefont {K.-L.~R.}\ \bibnamefont {Lo}},\ and\ \bibinfo {author} {\bibfnamefont {L.}~\bibnamefont {Sun}},\ }\bibfield  {title} {\bibinfo {title} {{Instability of Exotic Compact Objects and Its Implications for Gravitational-Wave Echoes}},\ }\href@noop {} {\  (\bibinfo {year} {2019})},\ \Eprint {https://arxiv.org/abs/1902.08180} {arXiv:1902.08180 [gr-qc]} \BibitemShut {NoStop}%
\bibitem [{\citenamefont {{Press}}(1979)}]{Press1979}%
  \BibitemOpen
  \bibfield  {author} {\bibinfo {author} {\bibfnamefont {W.~H.}\ \bibnamefont {{Press}}},\ }\bibfield  {title} {\bibinfo {title} {{On gravitational conductors, waveguides, and circuits.}},\ }\href {https://doi.org/10.1007/BF00756582} {\bibfield  {journal} {\bibinfo  {journal} {General Relativity and Gravitation}\ }\textbf {\bibinfo {volume} {11}},\ \bibinfo {pages} {105} (\bibinfo {year} {1979})}\BibitemShut {NoStop}%
\bibitem [{\citenamefont {Bemfica}\ \emph {et~al.}(2022)\citenamefont {Bemfica}, \citenamefont {Disconzi},\ and\ \citenamefont {Noronha}}]{Bemficaetal2020}%
  \BibitemOpen
  \bibfield  {author} {\bibinfo {author} {\bibfnamefont {F.~S.}\ \bibnamefont {Bemfica}}, \bibinfo {author} {\bibfnamefont {M.~M.}\ \bibnamefont {Disconzi}},\ and\ \bibinfo {author} {\bibfnamefont {J.}~\bibnamefont {Noronha}},\ }\bibfield  {title} {\bibinfo {title} {{First-Order General-Relativistic Viscous Fluid Dynamics}},\ }\href {https://doi.org/10.1103/PhysRevX.12.021044} {\bibfield  {journal} {\bibinfo  {journal} {Phys. Rev. X}\ }\textbf {\bibinfo {volume} {12}},\ \bibinfo {pages} {021044} (\bibinfo {year} {2022})},\ \Eprint {https://arxiv.org/abs/2009.11388} {arXiv:2009.11388 [gr-qc]} \BibitemShut {NoStop}%
\bibitem [{\citenamefont {Carr}(1975)}]{Carr:1975qj}%
  \BibitemOpen
  \bibfield  {author} {\bibinfo {author} {\bibfnamefont {B.~J.}\ \bibnamefont {Carr}},\ }\bibfield  {title} {\bibinfo {title} {{The Primordial black hole mass spectrum}},\ }\href {https://doi.org/10.1086/153853} {\bibfield  {journal} {\bibinfo  {journal} {Astrophys. J.}\ }\textbf {\bibinfo {volume} {201}},\ \bibinfo {pages} {1} (\bibinfo {year} {1975})}\BibitemShut {NoStop}%
\bibitem [{\citenamefont {Arbey}(2024)}]{Arbey:2024ujg}%
  \BibitemOpen
  \bibfield  {author} {\bibinfo {author} {\bibfnamefont {A.}~\bibnamefont {Arbey}},\ }\bibfield  {title} {\bibinfo {title} {{Primordial black holes, a small review}},\ }in\ \href@noop {} {\emph {\bibinfo {booktitle} {{58th Rencontres de Moriond on Very High Energy Phenomena in the Universe}}}}\ (\bibinfo {year} {2024})\ \Eprint {https://arxiv.org/abs/2405.08624} {arXiv:2405.08624 [gr-qc]} \BibitemShut {NoStop}%
\bibitem [{\citenamefont {Cacciapaglia}\ \emph {et~al.}(2024)\citenamefont {Cacciapaglia}, \citenamefont {Hohenegger},\ and\ \citenamefont {Sannino}}]{Cacciapaglia:2024wtp}%
  \BibitemOpen
  \bibfield  {author} {\bibinfo {author} {\bibfnamefont {G.}~\bibnamefont {Cacciapaglia}}, \bibinfo {author} {\bibfnamefont {S.}~\bibnamefont {Hohenegger}},\ and\ \bibinfo {author} {\bibfnamefont {F.}~\bibnamefont {Sannino}},\ }\bibfield  {title} {\bibinfo {title} {{Measuring Hawking Radiation from Black Hole Morsels in Astrophysical Black Hole Mergers}},\ }\href@noop {} {\  (\bibinfo {year} {2024})},\ \Eprint {https://arxiv.org/abs/2405.12880} {arXiv:2405.12880 [astro-ph.HE]} \BibitemShut {NoStop}%
\bibitem [{\citenamefont {Chitishvili}\ \emph {et~al.}(2023)\citenamefont {Chitishvili}, \citenamefont {Gogberashvili}, \citenamefont {Konoplich},\ and\ \citenamefont {Sakharov}}]{Chitishvili:2021jrx}%
  \BibitemOpen
  \bibfield  {author} {\bibinfo {author} {\bibfnamefont {M.}~\bibnamefont {Chitishvili}}, \bibinfo {author} {\bibfnamefont {M.}~\bibnamefont {Gogberashvili}}, \bibinfo {author} {\bibfnamefont {R.}~\bibnamefont {Konoplich}},\ and\ \bibinfo {author} {\bibfnamefont {A.~S.}\ \bibnamefont {Sakharov}},\ }\bibfield  {title} {\bibinfo {title} {{Higgs Field-Induced Triboluminescence in Binary Black Hole Mergers}},\ }\href {https://doi.org/10.3390/universe9070301} {\bibfield  {journal} {\bibinfo  {journal} {Universe}\ }\textbf {\bibinfo {volume} {9}},\ \bibinfo {pages} {301} (\bibinfo {year} {2023})},\ \Eprint {https://arxiv.org/abs/2111.07178} {arXiv:2111.07178 [astro-ph.HE]} \BibitemShut {NoStop}%
\bibitem [{\citenamefont {Andersson}\ and\ \citenamefont {Kokkotas}(1998)}]{Andersson:1997rn}%
  \BibitemOpen
  \bibfield  {author} {\bibinfo {author} {\bibfnamefont {N.}~\bibnamefont {Andersson}}\ and\ \bibinfo {author} {\bibfnamefont {K.~D.}\ \bibnamefont {Kokkotas}},\ }\bibfield  {title} {\bibinfo {title} {{Towards gravitational wave asteroseismology}},\ }\href {https://doi.org/10.1046/j.1365-8711.1998.01840.x} {\bibfield  {journal} {\bibinfo  {journal} {Mon. Not. Roy. Astron. Soc.}\ }\textbf {\bibinfo {volume} {299}},\ \bibinfo {pages} {1059} (\bibinfo {year} {1998})},\ \Eprint {https://arxiv.org/abs/gr-qc/9711088} {arXiv:gr-qc/9711088} \BibitemShut {NoStop}%
\end{thebibliography}%
\appendix
\end{document}